\documentclass{aa}

\usepackage{graphicx}
\usepackage[varg]{txfonts}
\usepackage[breaklinks=True]{hyperref}

\begin{document}

   \title{The fourth data release of the Kilo-Degree Survey:
   {\em ugri} imaging and nine-band optical-IR photometry over 1000 square degrees}
   
 \author{K. Kuijken\inst{1}
   \and C. Heymans\inst{2}
   \and A. Dvornik\inst{1}
   \and H. Hildebrandt\inst{3,4}
   \and J.T.A. de Jong\inst{5}
   \and A.H. Wright\inst{4}
   \and T. Erben\inst{4}
   \and M. Bilicki\inst{1,6}
   \and B. Giblin\inst{2}
   \and H.-Y. Shan\inst{4,7}
   \and F. Getman \inst{8}
   \and A. Grado\inst{8}
   \and H. Hoekstra \inst{1}
   \and L. Miller\inst{9}
   \and N. Napolitano\inst{10,8}
   \and M. Paolilo\inst{11}
   \and M. Radovich\inst{12}
   \and P. Schneider\inst{4}
   \and W. Sutherland\inst{13}
   \and M. Tewes\inst{4}
   \and C. Tortora\inst{14}
   \and E.A. Valentijn\inst{5}	
   \and G.A. Verdoes Kleijn\inst{5}
            }
   
   \institute{Leiden Observatory, Leiden University, P.O. Box 9513, 2300 RA Leiden, the Netherlands\\
              \email{kuijken@strw.leidenuniv.nl}
         \and
             Scottish Universities Physics Alliance, Institute for Astronomy, University of Edinburgh, Royal Observatory, Blackford Hill, Edinburgh, EH9 3HJ, UK
         \and
              Astronomisches Institut, Ruhr-Universit\"at Bochum, Universit\"atsstrasse 150, 44801 Bochum, Germany
         \and
              Argelander-Institut f\"ur Astronomie, Auf dem H\"ugel 71, D-53121 Bonn, Germany
         \and
              Kapteyn Astronomical Institute, University of Groningen, P.O. Box 800, 9700 AV Groningen, the Netherlands
         \and 
              Center for Theoretical Physics, Polish Academy of Sciences, al. Lotnik\'ow 32/46, 02-668, Warsaw, Poland
         \and 
              Shanghai Astronomical Observatory, Nandan Road 80, Shanghai 200030, China
         \and
              INAF - Osservatorio Astronomico di Capodimonte, Via Moiariello 16, 80131 Napoli, Italy
         \and
              Department of Physics, University of Oxford, Denys Wilkinson Building, Keble Road, Oxford OX1 3RH, UK
         \and
              School of Physics and Astronomy, Sun Yat-sen University, Guangzhou 519082, Zhuhai Campus, P.R. China 
         \and
              Dept.of Physics ``Ettore Pancini'', Universit\`a Federico II, 80126 Napoli, Italy
         \and
              INAF - Osservatorio Astronomico di Padova, via dell'Osservatorio 5, 35122 Padova, Italy
         \and
              School of Physics and Astronomy, Queen Mary University of London, Mile End Road, London E1 4NS, UK
         \and
              INAF - Osservatorio Astrofisico di Arcetri, Largo Enrico Fermi 5, 50125 Firenze, Italy
             }
   \date{Received ???; accepted ???}

  \abstract {The Kilo-Degree Survey (KiDS) is an ongoing optical
     wide-field imaging survey with the OmegaCAM camera at the VLT
     Survey Telescope, specifically designed for measuring weak
     gravitational lensing by galaxies and large-scale structure. When
     completed it will consist of 1350 square degrees imaged in four
     filters (\emph{ugri}). }
   {Here we present the fourth public data release which more than
     doubles the area of sky covered by data release 3. We also
     include aperture-matched \emph{ZYJHK$_{\rm s}$} photometry from our partner
     VIKING survey on the VISTA telescope in the photometry
     catalogue. We illustrate the data quality and describe the
     catalogue content.}
   {Two dedicated pipelines are used for the production of the optical
     data. The \textsc{Astro-WISE} information system is used for the
     production of co-added images in the four survey bands, while a
     separate reduction of the \emph{r}-band images using the {\sc theli}
     pipeline is used to provide a source catalogue suitable for the
     core weak lensing science case. 
     All data have been re-reduced for this data release using the latest versions of the pipelines. 
     The VIKING photometry is obtained
     as forced photometry on the {\sc theli} sources, using a re-reduction
     of the VIKING data that starts from the VISTA
     pawprints. Modifications to the pipelines with respect to earlier
     releases are described in detail.  The photometry is calibrated
     to the Gaia DR2 \emph{G} band using stellar locus regression.  }
   {In this data release a total of 1006 square-degree survey tiles
     with stacked \emph{ugri} images are made available, accompanied by
     weight maps, masks, and single-band source lists. We also provide
     a multi-band catalogue based on \emph{r}-band detections, including
     homogenized photometry and photometric redshifts, for the whole
     dataset.  Mean limiting magnitudes (5$\sigma$ in a 2\arcsec\
     aperture) and the tile-to-tile rms scatter are $24.23\pm0.12$,
     $25.12\pm0.14$, $25.02\pm0.13$, $23.68\pm0.27$ in \emph{ugri},
     respectively, and the mean \emph{r}-band seeing is 0\farcs70.  }
   {}


  \titlerunning{KiDS data release 4}
  \authorrunning{K. Kuijken et al.}

   \maketitle

\defcitealias{dejong/etal:2015}{[DR1/2]}
\defcitealias{dejong/etal:2017}{[DR3]}
\defcitealias{hildebrandt/etal:2017}{[KiDS450]}

\section{Introduction: the Kilo-Degree and VIKING Surveys}
\label{sec:intro}

High-fidelity images of the sky are one of the most fundamental kinds of data for astronomy research. While for many decades photographic plates dominated optical sky surveys, the advent of large-format CCD detectors for astronomy opened up the era of digital, high resolution, high sensitivity, linear-response images.

\begin{table*}
\caption{KiDS observing strategy: observing condition constraints and exposure times.}
\label{tab:ObservingConstraints}
\centering
\begin{tabular}{l c c c c c c c}
\hline\hline
Filter & Max. lunar & Min. moon & Max. seeing & Max. airmass & Sky transp. & Dithers & Total Exp.\\
~ & illumination & distance [deg] & [arcsec] & ~ & ~ & ~ & time [s] \\
\hline
\emph{u} & 0.4 & 90 & 1.1 & 1.2 & CLEAR & 4 & 1000 \\
\emph{g} & 0.4 & 80 & 0.9 & 1.6 & CLEAR & 5 & 900 \\
\emph{r} & 0.4 & 60 & 0.8 & 1.3 & CLEAR & 5 & 1800 \\
\emph{i} & 1.0 & 60 & 1.1 & 2.0 & CLEAR & 5 & 1200 \\
\hline
\end{tabular}
\end{table*}

The ESO VLT Survey Telescope \cite[VST;][]{capaccioli/schipani:2011,capaccioli/etal:2012} at ESO's Paranal observatory was specifically designed for wide-field, optical imaging. Its focal plane contains the square 268-million pixel CCD mosaic camera OmegaCAM \citep{kuijken:2011} that covers a $1\fdg013\times1\fdg020$ area at 0\farcs213 pitch, and the site and telescope optics (with actively controlled primary and secondary mirrors) ensure an image quality that is sub-arcsecond most of the time, and that does not degrade towards the corners of the field. Since starting operations in October 2011, more than half of the available time on the telescope has been used for a set of three wide-area `Public Imaging Surveys' for the ESO community. The Kilo-Degree Survey \cite[KiDS;][]{dejong/etal:2013}\footnote{\url{http://kids.strw.leidenuniv.nl}} is the deepest of these, and the one that exploits the best observing conditions. 

KiDS was designed as a cosmology survey, to study the galaxy population out to redshift $\sim$1 and in particular to measure the effect on galaxy shapes due to weak gravitational lensing by structure along the line of sight. By combining galaxy shapes with photometric redshift estimates it is possible to locate the redshift at which the gravitational lensing signal originates, and hence to map out the growth of large-scale structure, an important aspect of the evolution of the Universe and a key cosmology probe. Together with KiDS, two other major surveys are engaged in such measurements: the Dark Energy Survey \cite[DES;][]{des:2005}\footnote{\url{http://darkenergysurvey.org}} and the HyperSuprimeCam survey \cite[HSC;][]{aihara/etal:2018HSCdesc}\footnote{\url{http://hsc.mtk.nao.ac.jp/ssp/}}, and all three have reported intermediate cosmology results (\citealt{hildebrandt/etal:2017}, henceforth \citetalias{hildebrandt/etal:2017}; \citealt{troxel/etal:2018des,hikage/etal:2019}). Their precision is already such that the measurements can constrain some parameters in the cosmological model to a level that is comparable to what is achieved from the cosmic microwave background anisotropies \citep{planckVI:prep}. Since ground-based surveys are limited fundamentally by the atmospheric disturbance on galaxy shapes and photometry, space missions Euclid  
\citep{laureijs/etal:2011} and later WFIRST \citep{wfirst} are planned to increase the fidelity of such studies further.

   \begin{figure*}
   \centering
   \includegraphics[width=\textwidth]{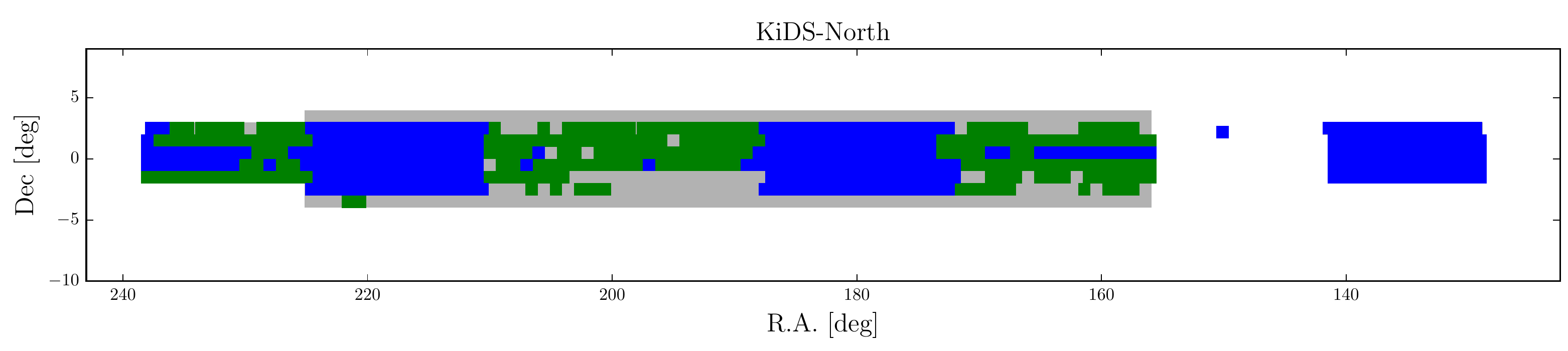}
   \includegraphics[width=\textwidth]{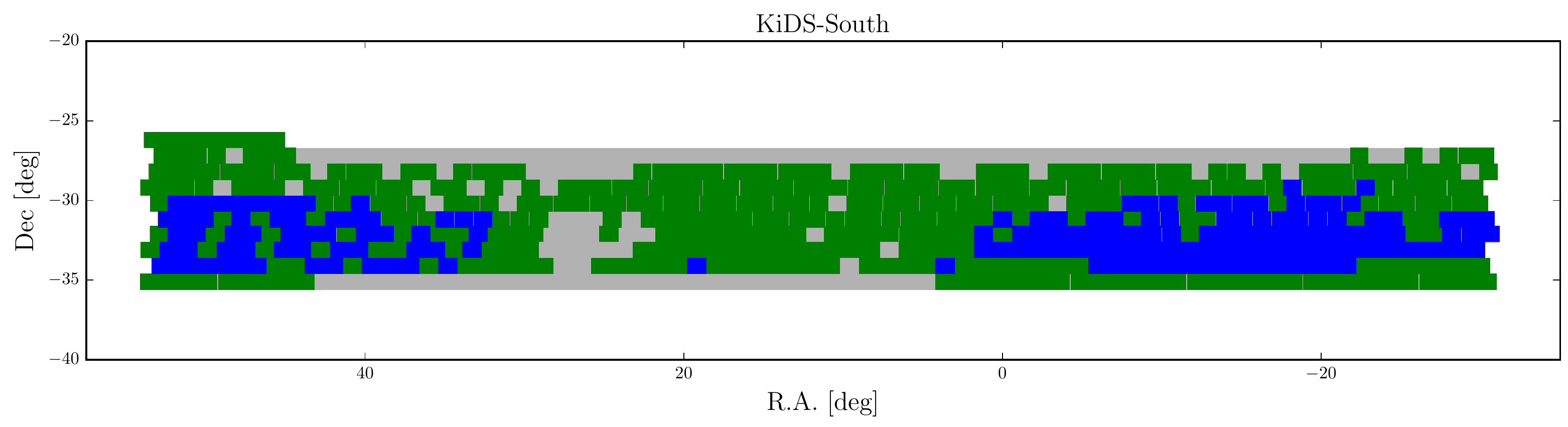}
   \caption{Sky distribution of survey tiles released in KiDS-ESO-DR4. Tiles shown in green are released for the first time; those in blue were included in the earlier data releases (DR1+2+3) but have been reprocessed for DR4. The full KiDS+VIKING area ($\sim1350 \deg^2$) is shown in grey. 
     {\it Top:} KiDS-North. {\it Bottom:} KiDS-South. The single
     pointing at RA=$150\deg$ is centred on the COSMOS/CFHTLS D2 field.}
              \label{fig:fields}
    \end{figure*}

To meet its primary science goal KiDS observes the sky in four bands: \emph{u}, \emph{g}, \emph{r} and \emph{i}. The \emph{r} band is used in dark time during the best seeing conditions (FWHM $<0\farcs8$), to make deep images for the measurement of galaxy shapes. In order to provide colours for photometric redshift estimates of the same sources, the \emph{r}-band data are supplemented with \emph{g}- and \emph{u}-band data taken in dark time of progressively worse seeing conditions ($<0\farcs9$ and $<1\farcs1$, respectively), and with \emph{i}-band data taken in grey or bright moon time with a mild seeing constraint ($<1\farcs1$). All observations consist of multiple dithered exposures to minimize the effect of gaps between the CCD's in the mosaic. Observing constraints and exposure times are summarized in Table~\ref{tab:ObservingConstraints}.

\begin{table}
\caption{Appriximate boundaries of the KiDS fields (see also Fig.~\ref{fig:fields}).}
\label{tab:fields}
\centering
\begin{tabular}{lll}
\hline\hline
Field & RA range & Dec range  \\
\hline\noalign{\smallskip}
KiDS-S & $[330\fdg0,52\fdg5]$ & $[-35\fdg6 , -26\fdg6]$ \\
\noalign{\smallskip}\hline\noalign{\smallskip}
KiDS-N  & $[155\fdg5,225\fdg5]$ & $[-4\fdg0 , +4\fdg0]$ \\
~         & $[225\fdg5,238\fdg5]$ & $[-2\fdg0 , +3\fdg0]$ \\
\noalign{\smallskip}\hline\noalign{\smallskip}
KiDS-N-W2 & $[128\fdg5,141\fdg5]$ & $[-2\fdg0 , +3\fdg0]$ \\
\noalign{\smallskip}\hline\noalign{\smallskip}
KiDS-N-D2 & $[149\fdg5,150\fdg5]$ & $[+1\fdg7 , +2\fdg7]$ \\
\hline
\end{tabular}
\end{table}

KiDS is targeting around 1350 square degrees of extragalactic sky, in two patches to ensure year-round observability. The Northern patch, KiDS-N, contains two additional smaller areas: KiDS-N-W2, which coincides with the G9 patch of the GAMA survey \citep{driver/etal:2011}, and KiDS-N-D2, a single pointing on the COSMOS field.  In a coordinated effort over the same part of the sky, the VISTA Kilo-degree INfrared Galaxy survey \cite[VIKING;][]{edge/etal:2013} on the nearby VISTA telescope added the five bands \emph{Z}, \emph{Y}, \emph{J}, \emph{H} and \emph{K}$_{\rm s}$. VIKING observations are complete\footnote{The originally planned KiDS area was 1500 square degrees, but this was reduced to match the footprint of the VIKING area.} and available in the ESO archive\footnote{\url{http://archive.eso.org}}.
Table~\ref{tab:fields} and Fig.~\ref{fig:fields} show the full KiDS footprint on the sky, as well as the part that is covered by the data contained in this data release (KiDS-ESO-DR4, or DR4 for short). The fields that were previously released under DR1+2+3 are also indicated: this is the area that was used for the \citetalias{hildebrandt/etal:2017} cosmic shear analysis, with the corresponding shape/photometric redshift catalogue released as DR3.1.

In order to improve the fidelity of the photometric redshift-based tomography, and to enable inclusion of high-value sources in the redshift range 0.9--1.2, \cite{hildebrandt/etal:prep} added VIKING photometry to the KiDS-450 data set, as described in \cite{wright/etal:prep}. The resulting cosmological parameter constraints of this new analysis, dubbed `KV450', are fully consistent with \citetalias{hildebrandt/etal:2017}. DR4 incorporates the methodology developed for KV450 and includes VIKING photometry for all sources.
Of all wide-area surveys, this makes it the one with by far the deepest near-IR data.

Though designed for the primary cosmology science case (\citetalias{hildebrandt/etal:2017}; \citealt{joudaki/etal:2017kids,vanuitert/etal:2018,koehlinger/etal:2017,harnoisderaps/etal:2017,amon/etal:2018EG,shan/etal:2018,martinet/etal:2018,giblin/etal:2018,asgari/etal:prep}), KiDS data are also being used for a variety of other studies, including the galaxy-halo connection \citep{vanuitert/etal:2016,vanuitert/etal:2017}, searches for strongly lensed galaxies \citep{petrillo/etal:2019} and quasars \citep{spiniello/etal:2018,sergeyev/etal:2018}, solar system objects \citep{mahlke/etal:2018}, photometric redshift machine learning method development \citep{amaro/etal:2019,bilicki/etal:2018}, studies of galaxy evolution  \citep{tortora/etal:2018DMfractions,tortora/etal:2018UCMG,roy/etal:2018}, bias \citep{dvornik/etal:2018}, environment \citep{brouwer/etal:2016,brouwer/etal:2018,costaduarte/etal:2018} and morphology \citep{kelvin/etal:2018}, galaxy group properties \citep{viola/etal:2015,jakobs/etal:2018}, galaxy cluster searches \citep{maturi/etal:2019,bellagamba/etal:2019}, intrinsic alignment of galaxies \citep{georgiou/etal:2019,johnston/etal:prep},  satellite halo masses \citep{sifon/etal:2015}, and searches for luminous red galaxies \citep{vakili/etal:prep} and quasars \citep{nakoneczny/etal:prep}.

The outline of this paper is as follows. Sect.~\ref{sec:dr4} is a discussion of the contents of KiDS-ESO-DR4. Sect.~\ref{sec:processing} summarises the differences in terms of processing and data products with respect to earlier releases. Sects.~\ref{sec:singlebandcats} and \ref{sec:ninebandcat} describe the single-band data products and the KiDS+VIKING nine-band catalogue, respectively. Sect.~\ref{sec:quality} illustrates the data quality. Data access routes are summarised in Sect.~\ref{sec:access} and a summary and outlook towards future data releases is provided in Sect.~\ref{sec:summary}. The Appendix gives a full listing of the information included in the images and catalogues, and of the data structure.

\begin{figure}
\includegraphics[width=0.5\textwidth,bb=0.3in 0 9.7in 4in]{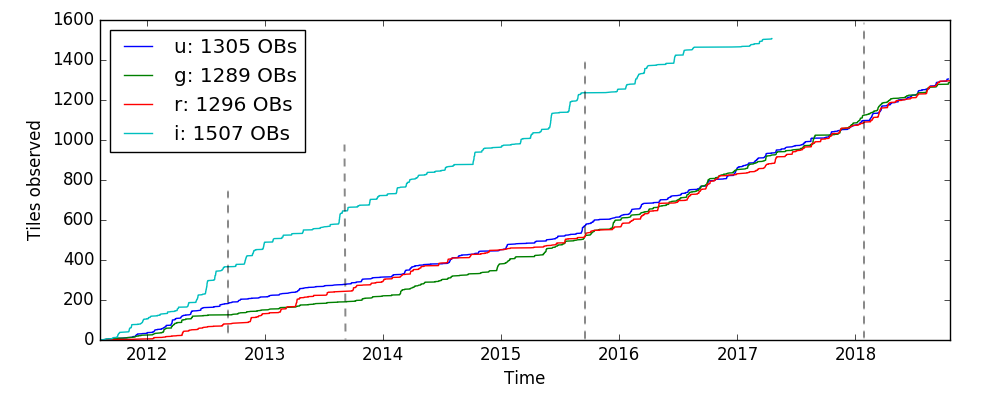}
\caption{Progress of the KiDS observations at the VST, in the four survey bands. Each Observing Block (OB) produces a square-degree co-added image. The \emph{i}-band data, for which data taking was significantly faster because of less competition for bright time on the telescope, had covered the originally planned 1500 square degree footprint by the time it was decided to limit the survey to the 1350 square degree area that comprise the completed VIKING area. The dashed lines indicate the cutoff dates for KiDS-ESO data releases 1 to 4.}
\label{fig:surveyprogress}
\end{figure}

\begin{figure*}
\includegraphics[width=\textwidth]{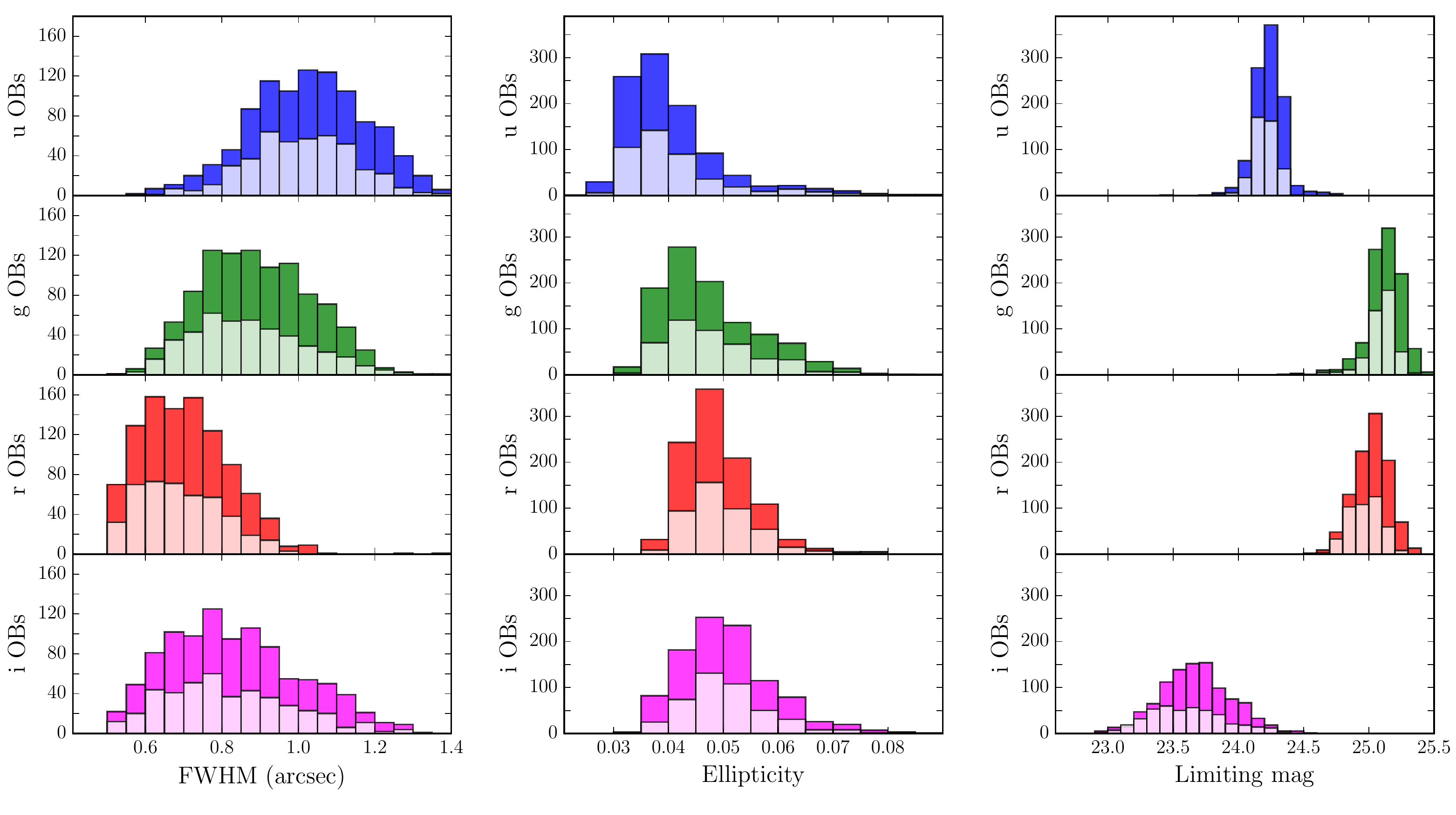}
\caption{Distributions of tile-by-tile data quality parameters for the KiDS DR4 data, grouped by filter, from top to bottom \emph{u}, \emph{g}, \emph{r} and \emph{i}. The light-coloured histograms represent the subset of the data that was previously released in DR1+2+3. Left: seeing. The differences between the bands reflect the observing strategy of reserving the best-seeing dark time for \emph{r}-band observations. Middle column: Average PSF ellipticity $\langle |e_{\rm psf}|\rangle$, where $e$ is defined as $1-b/a$ for major/minor axis lengths $a$ and $b$. Right: Limiting AB magnitude (5-$\sigma$ in a 2\arcsec\ aperture). The wider distribution of the \emph{i}-band observations is a caused by variations in the moon illumination, since the \emph{i}-band data were mostly taken in bright time. 
}
\label{fig:dataquality}
\end{figure*}

\section{The fourth KiDS data release}
\label{sec:dr4}

Unlike the previous incremental KiDS data releases, KiDS-ESO-DR4 represents a complete re-reduction of all the data using improved pipeline recipes and procedures. The differences will be described below. In terms of content of the data release, the main changes with respect to the earlier releases \citep{dejong/etal:2015,dejong/etal:2017} are the more than doubling of the area, and the inclusion of photometry from the near-IR VIKING images into the multi-band catalogue. Whereas the sky coverage of the earlier data releases was still quite fragmented, DR4's greater homogeneity will for the first time enable wide-area studies over the full length of the survey patches.

KiDS observations consist of individual square-degree tiles. Each tile is covered by a set of five (four in \emph{u}) dithered exposures with OmegaCAM/VST, consisting of 32 individual CCD images each. The dither step sizes are matched to the gaps between CCD's (25\arcsec\ in RA, 85\arcsec\ in declination), to ensure that each part of the tile is covered by at least three (two in \emph{u}) sub-exposures.  Overlaps between adjacent tiles are small, of order 5\%. All observations in a single band are taken in immediate succession (KiDS is not designed for variability measurements), but there is no constraint on the time between observations of any given tile in the different filters. Typically the shutter is closed for 35-60 seconds between the sub-exposures to allow for CCD readout, telescope repointing and active optics adjustments.

With little exception, the DR4 data comprise all KiDS tiles for which the 4-band observations had been taken by January 24th, 2018. Over half of the data is from after mid-2015, which is when the VST saw several improvements that affect the quality of the data (and improved operational efficiency as well, see Fig.~\ref{fig:surveyprogress}). The two main improvements were (i) the baffling of the telescope was improved to the point that stray light from sources outside the field of view of the camera was drastically reduced; and (ii) the on-line image analysis system (based on simultaneous pre- and post-focus star images at the edge of the field of view) was modified to control also the tilt of the secondary mirror, improving pointing and especially off-axis image quality.

Figure~\ref{fig:dataquality} shows the distributions of key data quality parameters of the observations: point spread function (PSF) full width at half maximum (FWHM), average PSF ellipticity and limiting magnitude. It illustrates that the global quality of the DR4 data is very similar to the earlier KiDS data releases. Limiting AB magnitudes (5-$\sigma$ in 2\arcsec aperture) are $24.23\pm0.12$, $25.12\pm0.14$, $25.02\pm0.13$, $23.68\pm0.27$ in \emph{ugri}, respectively, with the error bars representing the RMS scatter from tile to tile. The mean seeing in \emph{r} band is $0\farcs70$.

\begin{figure}
\includegraphics[width=0.5\textwidth,bb=220 10 1000 730]{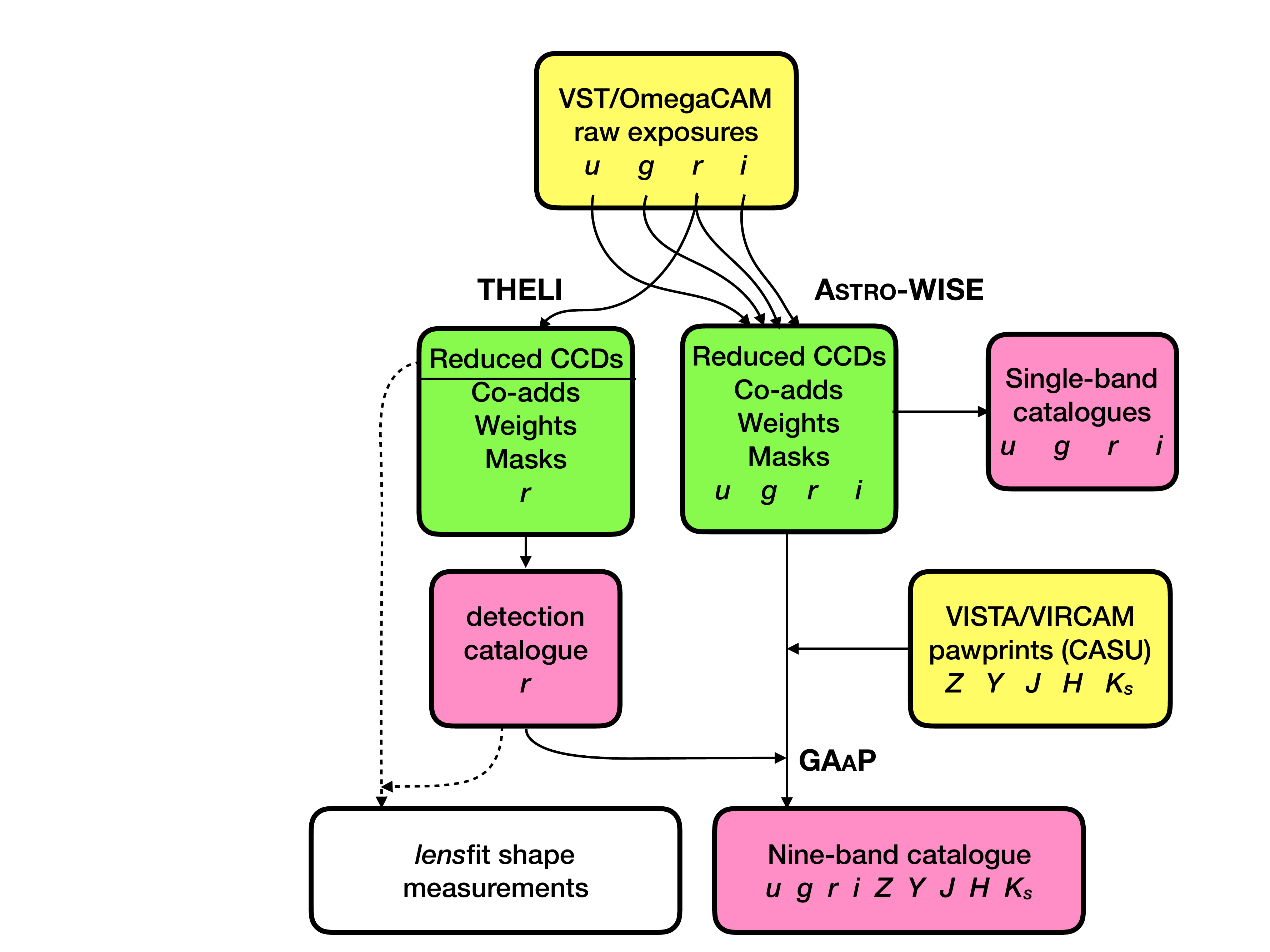}
\caption{\label{fig:DR4flowchart} Schematic representation of the DR4 processing steps and content. Yellow boxes show the input data from VST and VISTA, green indicates image products, and source lists are shown in pink. The \emph{lens}fit-based lensing measurements -- initially not released in DR4 -- are shown as dotted lines.}
\end{figure}

The KiDS images were processed with two independent pipelines, as was the case for the KiDS-450 weak lensing analysis that was based on KiDS-ESO-DR3. The {\sc Astro-WISE} pipeline and data reduction environment \citep{mcfarland/etal:2013}\footnote{\url{http://www.astro-wise.org}} was used to produce stacked images in the four bands, from which the photometry in the catalogues is obtained. The {\sc theli} pipeline \citep{erben/etal:2005}\footnote{\url{https://www.astro.uni-bonn.de/theli/}}, which is optimised for weak lensing measurements, was used for a separate reduction of the \emph{r}-band data only. In order to have consistent source lists, the detection and astrometry of the \emph{r}-band sources is performed on these {\sc theli} images (which are also the ones which are used for the weak lensing measurements). This detection catalogue is then used as the basis of list-driven `forced' photometry on the \emph{u}, \emph{g}, \emph{r} and \emph{i} {\sc Astro-WISE} stacked images and the VIKING \emph{Z}, \emph{Y}, \emph{J}, \emph{H} and \emph{K}$_{\rm s}$ images. \footnote{Note that for DR4 there is no separate multi-band catalogue based on \emph{r}-band detections in the {\sc Astro-WISE} data, as there was for \citetalias{dejong/etal:2017}.}
The data flow is summarized in Fig.~\ref{fig:DR4flowchart}. 

The set of KiDS-ESO-DR4 data products includes 5030 separate co-added images (1006 square-degree tiles in the \emph{ugri} filters, plus the separate \emph{r}-band co-adds from {\sc theli}), with corresponding weight and mask flag images. The images are photometrically and astrometrically calibrated using a combination of nightly photometric calibration information, the Gaia DR2 photometry \citep{brown/etal:2018} with stellar locus regression, and the SDSS and 2MASS astrometry \citep{alam/etal:2015,skrutskie/etal:2006}. Each image has a corresponding source catalogue as well. In addition, catalogues of nine-band \emph{ugriZYJHK}$_{\rm s}$ photometry are provided containing list-driven (i.e., forced), PSF- and aperture-matched photometry using the {\sc GAaP} technique \citep{kuijken/etal:2015}, applied to the KiDS tiles and the overlapping VIKING data.

\section{Data processing}
\label{sec:processing}
For details of the image processing pipelines we refer to the description in the DR1/2 and DR3 release papers \citep[henceforth \citetalias{dejong/etal:2015}]{dejong/etal:2015} and \citep[henceforth \citetalias{dejong/etal:2017}]{dejong/etal:2017}, noting the changes described in Sects.~\ref{sec:AWchanges} and \ref{sec:THELIchanges} below.

The DR4 catalogues are of two types: single-band catalogues for the four KiDS bands, and a combined nine-band catalogue that includes KiDS and VIKING photometry for the \emph{r}-band detected sources.

\subsection{Changes to the {\sc Astro-WISE} image processing pipeline}
\label{sec:AWchanges}

\subsubsection{Co-added image creation}
The production of KiDS-ESO-DR4 includes $4\times1006$ OmegaCAM tiles, which required
some 611,000 individual CCD exposures, as well as associated calibration observations, to be processed. The automatic processing steps followed closely those described in \citetalias{dejong/etal:2017}. Individual CCD exposures are corrected for electronic crosstalk, bias corrected, flat-fielded, illumination corrected and (for \emph{i} band only) fringe-subtracted. New coefficients were determined for the electronic crosstalk correction between CCD's \verb ESO_CCD_#95  and \verb ESO_CCD_#96  (see \citetalias{dejong/etal:2015}) with validity periods determined by maintenance or changes to the instrument; these are listed in Table~\ref{tab:xtalk}. An automatic mask is also generated for each CCD, which marks the location of saturated, hot, and cold pixels, as well as satellite tracks identified through a Hough transform analysis.

\begin{table}
\caption{Applied cross-talk coefficients.}
\label{tab:xtalk}
\centering
\footnotesize
\resizebox{\columnwidth}{!}{
\begin{tabular}{l | c c | c c}
\hline\hline
Period & \multicolumn{2}{c|}{CCD \#95 to CCD \#96\tablefootmark{a}} & \multicolumn{2}{c}{CCD \#96 to CCD \#95\tablefootmark{a}} \\
~ & $a$ & $b$ ($\times10^{-3}$) & $a$ & $b$ ($\times10^{-3}$)\\
\hline
2011-08-01 - 2011-09-17 & $-$210.1 & $-$2.504 & 59.44 & 0.274 \\
2011-09-17 - 2011-12-23 & $-$413.1 & $-$6.879 & 234.8 & 2.728 \\
2011-12-23 - 2012-01-05 & $-$268.0 & $-$5.153 & 154.3 & 1.225 \\
2012-01-05 - 2012-07-14 & $-$499.9 & $-$7.836 & 248.9 & 3.110 \\
2012-07-14 - 2012-11-24 & $-$450.9 & $-$6.932 & 220.7 & 2.534 \\
2012-11-24 - 2013-01-09 & $-$493.1 & $-$7.231 & 230.3 & 2.722 \\
2013-01-09 - 2013-01-31 & $-$554.2 & $-$7.520 & 211.9 & 2.609 \\
2013-01-31 - 2013-05-10 & $-$483.7 & $-$7.074 & 224.7 & 2.628 \\
2013-05-10 - 2013-06-24 & $-$479.1 & $-$6.979 & 221.1 & 2.638 \\
2013-06-24 - 2013-07-14 & $-$570.0 & $-$7.711 & 228.9 & 2.839 \\
2013-07-14 - 2014-01-01 & $-$535.6 & $-$7.498 & 218.9 & 2.701 \\
2014-01-01 - 2014-03-08 & $-$502.2 & $-$7.119 & 211.6 & 2.429 \\
2014-03-08 - 2014-04-12 & $-$565.8 & $-$7.518 & 215.1 & 2.578 \\
2014-04-12 - 2014-08-12 & $-$485.1 & $-$6.887 & 201.6 & 2.237 \\
2014-08-12 - 2014-01-09 & $-$557.9 & $-$7.508 & 204.2 & 2.304 \\
2014-01-09 - 2015-05-01 & $-$542.5 & $-$7.581 & 219.9 & 2.535 \\
2015-05-01 - 2015-07-25 & $-$439.3 & $-$6.954 & 221.5 & 2.395 \\
2015-07-25 - 2015-08-25 & $-$505.6 & $-$7.535 & 229.7 & 2.605 \\
2015-08-25 - 2015-11-10 & $-$475.2 & $-$7.399 & 218.0 & 2.445 \\
2015-11-10 - 2016-06-17 & $-$457.8 & $-$6.831 & 201.6 & 2.212 \\
2016-06-17 - 2016-06-25 & $-$351.8 & $-$4.973 &165.3 & 1.168 \\
2016-06-25 - 2016-09-08 & $-$476.3 & $-$6.920 & 200.4 & 2.202 \\
2016-09-08 - 2017-08-01 & $-$465.3 & $-$6.594 & 184.7 & 1.980 \\
2017-08-01 - 2018-02-15 & $-$492.3 & $-$6.480 & 169.9 & 1.802 \\
\hline
\end{tabular}
}
\tablefoot{
\tablefoottext{a}{
Correction factors $a$ and $b$ are applied to each pixel in the target CCD based on the pixel values in the source CCD:
\begin{equation}
I'_i =
\begin{cases}
I_i + a, &\text{if $I_j = I_{\rm{sat.}}$;}\\
I_i + b  I_j, &\text{if $I_j < I_{\rm{sat.}}$,}\\
\end{cases}
\end{equation}
where $I_i$ and $I_j$ are the pixel values in CCDs $i$ and $j$, $I'_i$ is the corrected pixel value in CCD $i$ due to cross-talk from CCD $j$, and $I_{\rm{sat.}}$ is the saturation pixel value.
}
}
\end{table}

These `reduced science frames' are then astrometrically calibrated and regridded using the {\sc scamp} and {\sc swarp} software \citep{bertin/etal:2002,bertin:2006,bertin:2010scamp,bertin:2010swarp}, in two steps: first a `local' step establishes a per-detector solution using the 2MASS stars in the frame\footnote{We have found the 2MASS catalogue to be sufficient as astrometric reference, but intend to move to Gaia in the future.}, and then these solutions are refined using {\sc scamp} into a tile-wide `global solution' for the full co-added stacked image that uses the information from fainter overlapping objects. For DR4 the astrometric solution was made more robust by starting from a model of the focal plane that accounts for detector array lay-out and instrument optics distortion, and using this as input to {\sc scamp}. Using {\sc swarp}, the global astrometric solution is then used to resample each CCD exposure into a `regridded science frame', with a uniform 0\farcs20 pixel grid with tangent projection centred on the nominal tile centre. During this step the background, determined by interpolating a $3\times3$ median-filtered map of background estimates in $128\times128$ pixel blocks, is subtracted. Finally these regridded images are co-added, taking account of the weight maps generated by {\sc swarp}, and the masks. Each co-added image is about $18,500\times19,500$ pixels in size, and takes up about 1.5 Gbyte of storage. For every co-added image a mask that flags reflection haloes of bright stars in the field is also produced, using the {\sc pulecenella} code developed for \citetalias{dejong/etal:2015} (see Sect.~\ref{sec:singlebandcats}).

$12\times12$ binned versions of the co-added images (at two contrast settings) and of the weight image are then visually inspected, together with a set of diagnostic plots that show the PSF ellipticity and size as function of position on the field, astrometry solution residuals, and the PSF size before and after co-addition. The main issues that get flagged at this stage are (i) residual satellite tracks (ii) background features associated with stray light casting shadows of the baffles mounted above the CCD bond wires (iii) unstable CCDs (gain jumps) (iv) residual fringing in the background of the \emph{i}-band images or (v) large-scale reflections. In DR3 any issues found at this stage were addressed by masking the co-added image, even though in many cases the problem only affected one sub-exposure. In DR4 issues (i)--(iii) were solved with new procedures, as described below. The other cases are still in the data: the residual fringes are not corrected for but will be fixed with re-observations of the full survey footprint in the \emph{i} band, and large-scale reflections need to be masked manually or otherwise identified in the catalogues as groups of sources with unusual colours.

The new procedures for removing residual satellite tracks, bond wire baffle features, and unstable CCDs, involve a minimum of manual intervention. The satellite tracks are marked by clicking on their ends on a display of the inspection JPG images, after which an automatic procedure converts the pixel positions to sky coordinates, checks which of the sub-exposures that contribute to the co-added image contains the track, measures the track's width, and updates the corresponding CCDs' masks before stacking anew. Similarly, the bond wire baffle features have a typical width and all the inspector needs to do is to indicate whether the shadow is visible on the upper, lower or both sides of the baffles, so that the corresponding lines in the CCD images can be masked. Unstable CCDs are simply removed from the list of exposures to be co-added.

\subsubsection{PSF Gaussianization and \textsc{GAaP} photometry}
\label{sec:gaap}
The Gaussian Aperture and PSF ({\sc GAaP}) photometry method \citep{kuijken/etal:2015} developed for KiDS multi-band photometry was improved further. {\sc GAaP} entails (i) convolving each image with a spatially variable kernel designed to render the PSF homogeneous and Gaussian,  (ii) defining a pre-seeing Gaussian elliptical aperture function for every source, and (iii)  for every band deconvolving this aperture by the corresponding Gaussianized PSF and performing aperture photometry. 
The method is superior to traditional techniques such as dual-image mode {\sc SExtractor} measurements as it explicitly allows for PSF differences between exposures in multiple bands, and it reduces noise by measuring colours from the highest SNR part of the sources. A demonstration of the improvement provided by {\sc GAaP} photometry is given in \cite{hildebrandt/etal:2012}. 
{\sc GAaP} photometry gives colours that are corrected for PSF differences, but when the source is more extended than the aperture function the fluxes are underestimates of the total flux. For stars and other unresolved sources {\sc GAaP} fluxes {\em are} total fluxes.

For DR4 we have modified the procedure for step (i). We still use the several thousand stars in each image as samples of the PSF, but rather than first modelling the PSF $P$ as a spatially varying, truncated shapelet expansion, from which the convolution kernel is then constructed in shapelet coefficient space, we now directly solve for the kernel shapelet coefficients that give a Gaussian PSF in pixel space. Thus we obtain the kernel coefficients $k_{abc}$ of the shapelet components $S_{ab}^{\beta_c}$ as the least-squares solution of
\begin{equation}
\label{eq:kernel}
\sum_{abc}k_{abc}\left[S_{ab}^{\beta_c}\otimes P\right](x,y)={\exp\left[{-\left(x^2+y^2\right)/2\beta_g^2}\right]\over2\pi\beta_g^2} .
\end{equation}
As in \citetalias{dejong/etal:2017}, the size $\beta_g$ of the target Gaussian is set to $1.3$ times the median dispersion of Gaussian fits to the stars in the image. The fit of Eq.\,(\ref{eq:kernel}) is performed over all pixels out to $15\beta_g$ from the centre of each star.
In our implementation we use terms with $\beta_1=\beta_g$, $a+b\le8$ plus a set of wider shapelets with $\beta_2=2.5\beta_g$, $3\le a+b\le6$ specifically designed to model the wings of the kernel better. The large-$\beta$ shapelets with $a+b<3$ are not included in the series as they are not sufficiently orthogonal to the small-$\beta$ terms.

\subsubsection{Photometric calibration using Gaia and stellar locus regression}
\label{sec:photcal}

\begin{figure}
\includegraphics[width=0.5\textwidth,bb=0.3 0 6in 4.3in,clip=true]{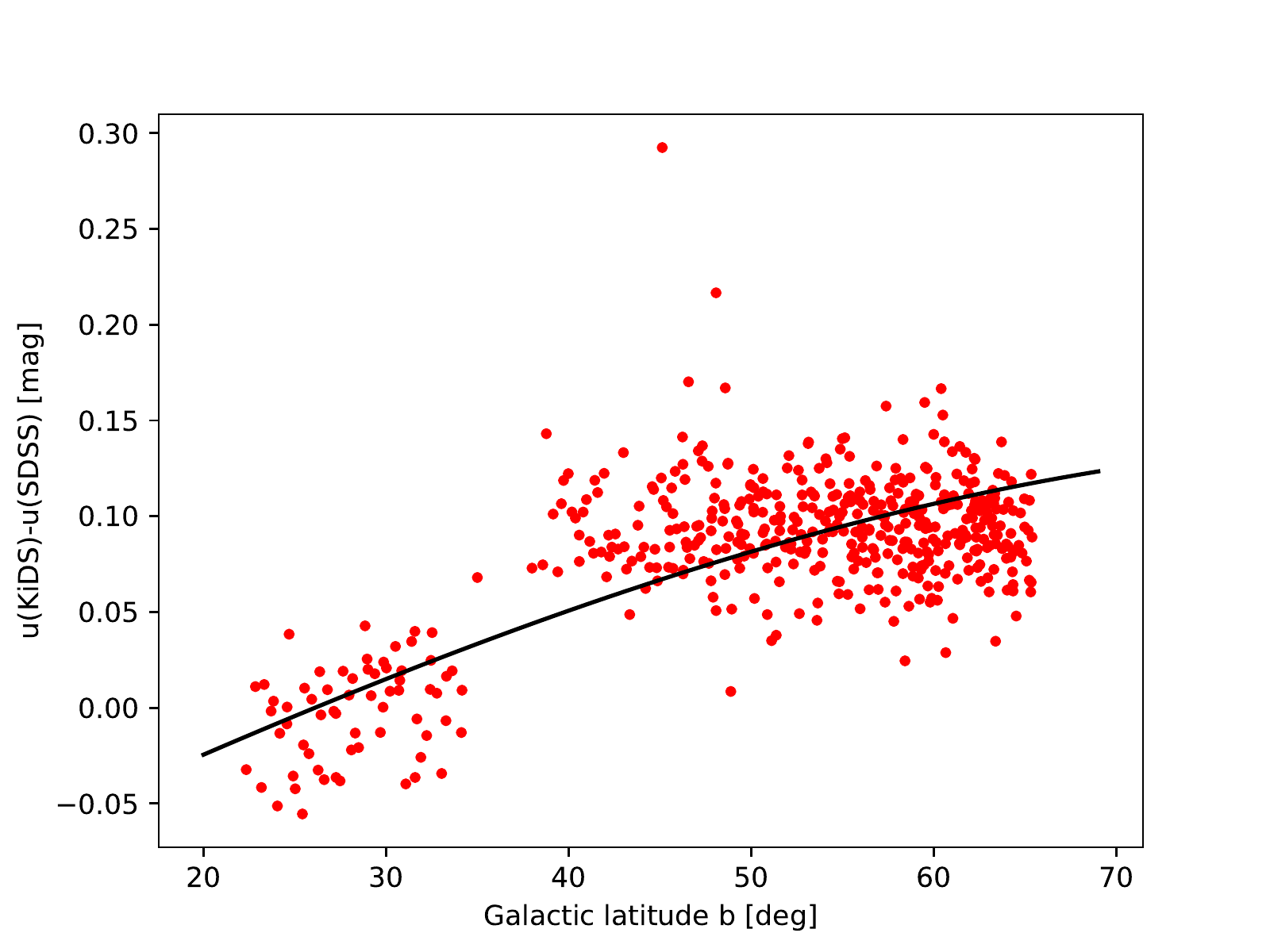}
\caption{Residual \emph{u}-band magnitude variation when Eq.\,(\ref{eq:SLRu}) is used without a dependence on Galactic latitude $b$ (i.e., $f(b)=0$). Each point shows the average offset $\hbox{\emph{u}}_{\rm KiDS}-\hbox{\emph{u}}_{\rm SDSS}$ for the calibration stars in a separate KiDS-N tile. The line represents the Galactic latitude correction of Eq.\,(\ref{eq:latcorr}).}
\label{fig:latitudecorrection}
\end{figure}

\begin{figure}
   \includegraphics[width=0.5\textwidth]{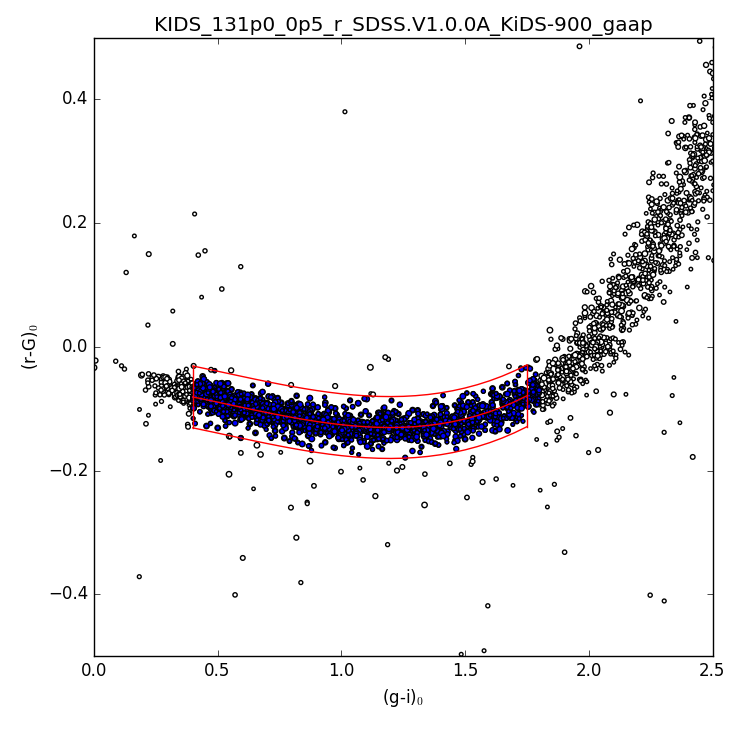}
\caption{Colour-colour relation used to calibrate the KiDS \emph{r}-band measurements to the Gaia \emph{G}-band catalogue, for an example tile. The blue points and the box indicate the dereddened $(g-i)$ colour range for the stars used,
and the $\pm0.05$ magnitude iterative clipping width about the fiducial sequence. The shape of the sequence is determined from the overlap area between KiDS-N and SDSS.}
\label{fig:gaiacalib}
\end{figure}

\begin{figure*}
\includegraphics[width=\textwidth]{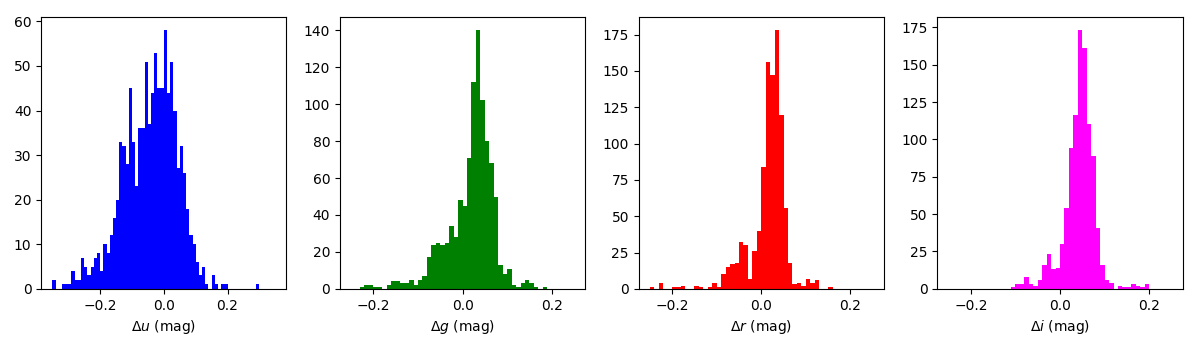}
\caption{Distribution of the \emph{u}, \emph{g}, \emph{r} and \emph{i} zeropoint corrections using the stellar locus regression plus Gaia calibration.}
\label{fig:zpts}
\end{figure*}

All KiDS `reduced science frame' images are initially put on a photometric scale by using nightly zeropoints derived from standard star observations taken in the middle of the night. For DR4, these zeropoints are refined with a combination of Stellar Locus Regression (SLR) and calibration to Gaia photometry. We use {\sc GAaP} photometry for the stars: this is appropriate since for unresolved sources the {\sc GAaP} magnitudes correspond to the total flux of the source.

With the advent of Gaia Data Release 2 \citep{brown/etal:2018} a deep, homogeneously calibrated, optical all-sky catalogue is now available. Each KiDS tile contains several thousand Gaia stars, with broad-band photometry measurements that are individually accurate to better than 0.01 magnitudes\footnote{Note that the Gaia DR2 \emph{g}-band calibration differs from what was used in Gaia DR1, through a new determination of the filter bandpass.}. KiDS DR4 photometry is calibrated to the Gaia DR2 catalogue in two steps. First, we calibrate the colours $u-g$, $g-r$ and $r-i$ by comparing the stellar colour-colour diagrams to fiducial sequences, using the `stellar locus regression' described in \citetalias{dejong/etal:2017}. Based on the work of  \cite{ivezic/etal:2004}, four principal colours $P2s$, $P2w$, $P2x$ and $P2k$ were initially used to derive colour offsets (see appendix B of \citetalias{hildebrandt/etal:2017}). After this procedure, while validating the results by comparing the magnitudes of stars in KiDS-N and SDSS, we found that the $P2k$ principal colour, which is the most sensitive to the \emph{u} band, gave unreliable results. The \emph{u}-band zeropoints were therefore determined using a modified procedure, which was found to be more robust, as follows. For stars in Gaia with dereddened KiDS {\sc GAaP} colour $(g-r)_0$ between 0.15 and 0.8 and $\hbox{\emph{u}}<21$, the quantity
\begin{equation}
\label{eq:SLRu}
  {\cal U}=g_0 + 2(g-r)_0 +0.346 + \max\left\lbrace 0, 2 \left[ 0.33-(g-r)_0\right]\right\rbrace - f(b)
\end{equation}
is a good predictor of the SDSS \emph{u} magnitude. The Galactic latitude dependence $f(b)$, shown in Fig.~\ref{fig:latitudecorrection}, can be fitted as
\begin{equation}
\label{eq:latcorr}
  f(b)=0.25\, |\sin b\,| - 0.11 .
\end{equation}
Stars closer to the Galactic plane are fainter in \emph{u} than the mean $(u-g,g-r)$ colour-colour relation predicts, as is expected if the high-latitude sample is dominated by halo stars of lower metallicity than the disk stars found at low $b$.
Since KiDS-N spans latitudes from $22\degr$ to $66\degr$, the correction is significant. KiDS-S covers latitudes from the South Galactic Pole to $-53\degr$, and though we cannot test the latitude dependence for this part of the survey because of a lack of a suitable calibration data set, we have applied the same correction.
We therefore adjust the KiDS \emph{u}-band zeropoints in each tile until the average ${\cal U}-u=0$, with an iterative clipping of outliers more than 0.1 mag from the mean relation. For applications where \emph{u}-band photometry is critical, we caution that because of the uncertain functional form of the latitude dependence, currently the calibration of this band is subject to a residual uncertainty of up to 0.05 magnitudes.

The zeropoint of the \emph{r}-band magnitude is then tied to Gaia by matching the dereddened $(r-G,g-i)$ relation to the one followed by the stars in the SDSS-KiDS overlap region (Fig.~\ref{fig:gaiacalib}). As reported in \citetalias{dejong/etal:2017}, there is a slight colour term between the SDSS and KiDS \emph{r} filters: we have arbitrarily forced the KiDS and SDSS \emph{r}-band zeropoints to agree for stars of $(g-i)_0=0.8$, adopting 
\begin{equation}
r_{\rm KiDS}-r_{\rm SDSS}=-0.02 [(g-i)_0-0.8].
\label{eq:rkidssdss}
\end{equation}

For DR4, extinction corrections were derived using the \cite{schlegel/etal:1998} $E(B-V)$ map in combination with the $R_V=3.1$ extinction coefficients from \cite{schlafly/finkbeiner:2011}\footnote{The earlier data releases used the \cite{schlegel/etal:1998} coefficients.}. For the \emph{ugri} filters we adopt the corresponding SDSS filter values. Since the VISTA bands were not included in the \cite{schlafly/finkbeiner:2011} tables, as an approximation we have taken the values for SDSS \emph{z}, LSST \emph{y}, and UKIRT \emph{JHK} filters.
From a regression of $r_{\rm SDSS}-G$ vs. $E(B-V)$ we derive the \emph{G}-band extinction coefficient as $A_G/A_r=0.96$. Given that $E(B-V)$ values in the KiDS tiles are typically below $0.05$ magnitudes, residual uncertainties in these coefficients are of little consequence. The adopted extinction coefficients are summarised in Table~\ref{tab:extinction}.

\begin{table}
\caption{\label{tab:extinction} Extinction coefficients $R_f=A_f/E(B-V)$ used in this work, from \cite{schlafly/finkbeiner:2011} (SF11). These coefficients are used to scale the $E(B-V)$ values in the \cite{schlegel/etal:1998} map. }
\centering
\begin{tabular}{crl}
\hline\hline
Filter $f$ & \multicolumn{1}{c}{$R_f$} & Source\\
\hline
\emph{u} &4.239 & SF11 (SDSS)\\
\emph{g} & 3.303 & SF11 (SDSS)\\
\emph{r} & 2.285 & SF11 (SDSS)\\
\emph{i} & 1.698 & SF11 (SDSS)\\
\emph{G} & 2.194 & This work\\
\emph{Z} & 1.263& SF11 (SDSS \emph{z})\\
\emph{Y} & 1.088& SF11 (LSST \emph{y})\\
\emph{J} & 0.709& SF11 (UKIRT) \\
\emph{H} & 0.449& SF11 (UKIRT)\\
\emph{K}$_{\rm s}$ &0.302 & SF11 (UKIRT)\\
\hline
\end{tabular}\\
\end{table}

This direct, tile-by-tile calibration of the KiDS photometry to Gaia obviates the need for the overlap photometry that was used in \citetalias{dejong/etal:2017}. Effectively, we are tying KiDS to the Gaia DR2 \emph{G}-band photometry, and anchoring it to the SDSS calibration.
Specifically, calibrating each KiDS tile to Gaia DR2 involves the following steps:
\begin{enumerate}
\item Select Gaia stars in the tile area with $16.5<G<20$, and with unflagged photometric measurements.
\item Keep those stars with SLR-calibrated, dereddened $(g-i)_0$ colours in the range $[0.4,1.8]$ .
\item Predict dereddened $(r-G)_0$ values from these $(g-i)_0$ colours using the following relation, obtained by fitting the difference between the predicted \emph{r}$_{\rm KiDS}$ (from Eq.\,\ref{eq:rkidssdss}) and the measured \emph{G} in the KiDS-SDSS overlap region:
\begin{equation}
(r_{\rm KiDS}-G)_0= -0.0618 - 0.0724 y + 0.0516 y^2 + 0.0665 y^3
\end{equation}
where $y=(g-i)_0-0.8$ (see Fig.~\ref{fig:gaiacalib}).
\item Determine the median offset between this fiducial $r-G$ and the measured value, using iterative clipping. 
\item Apply this median offset to the \emph{ugri} magnitudes for all sources in the tile.
\end{enumerate}
Figure~\ref{fig:zpts} shows the distribution of tile-by-tile zeropoint corrections that have been applied to the magnitudes in the catalogues. Typical values are of the order of 0.05--0.1 magnitudes, and about twice that in the \emph{u} band.

Note that the SLR procedure aligns the dereddened stellar loci of all the tiles, assuming that all the dust is in the foreground and not mixed in with the stars (which is a reasonable assumption given the high
Galactic latitude and bright-end limit of the calibration stars).

In DR4 all {\sc GAaP} photometry is performed twice, with minimum aperture settings MIN{\_}APER=0\farcs7 and 1\farcs0 (see Sect.~\ref{sec:ninebandcat} below). The photometric zeropoint determinations are also performed with both settings, and the one based on the largest number of stars with valid measurements is recorded in the header of the images and single-band source catalogues with the DMAG keyword. CALMINAP gives the value of MIN{\_}APER that was used for the calibration, and CALSTARS is the number of Gaia stars with valid measurements.

\subsection{Changes to the {\sc theli} pipeline}
\label{sec:THELIchanges}

In \citetalias{hildebrandt/etal:2017}, the {\sc theli} pipeline was used to process the \emph{r}-band images for the weak lensing analysis, which resulted in two different \emph{r}-band source catalogues with multi-band photometry (DR3 and the `lensing catalogue' DR3.1). For DR4 we unify the analysis, using the \emph{r}-band sources detected on the {\sc theli} co-added images as the basis of the multi-band photometry as well as the forthcoming lensing analyses.

\begin{figure}
\includegraphics[width=0.5\textwidth]{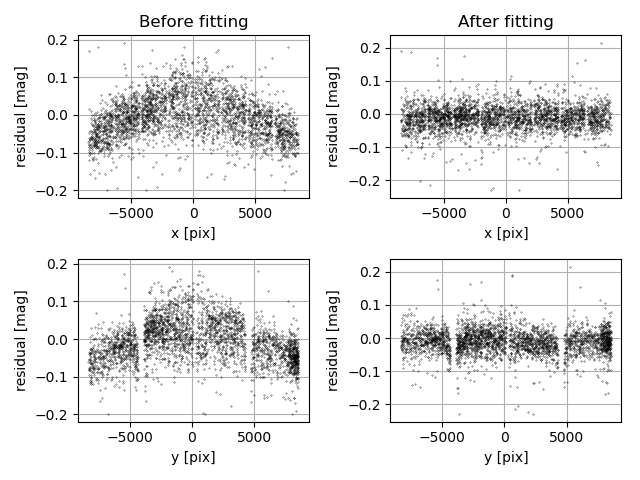}
\caption{Photometric illumination correction of the weak-lensing {\sc theli}
data: Flat-fielded OmegaCAM data show systematic zeropoint variations
over the field-of-view if the complete mosaic is calibrated with a
single photometric zeropoint (left panels). The residuals can well be
fitted and corrected with a two-dimensional, second-order polynomial
over the field-of-view (right panels)}
\label{fig:illum_theli}
\end{figure}

\begin{figure}
\includegraphics[width=0.5\textwidth]{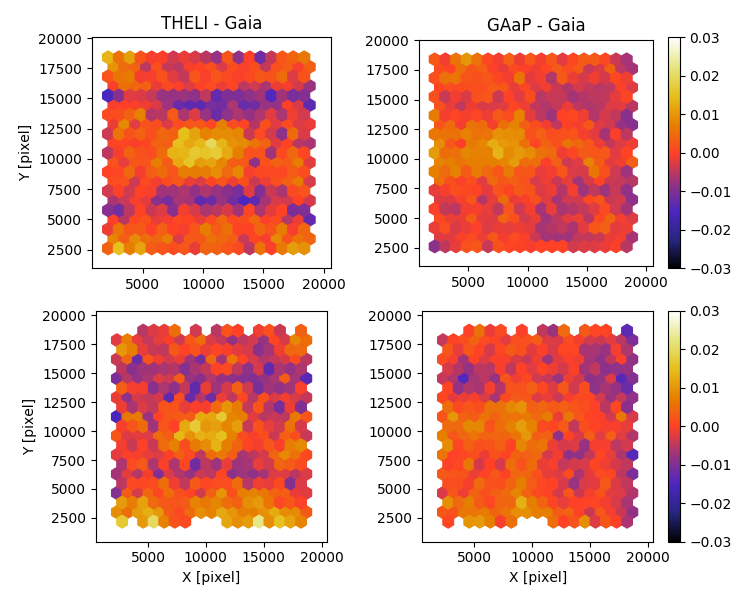}
\caption{Check of the illumination corrections, through direct comparison of star magnitudes with Gaia. Each panel shows the median difference between a KiDS-DR4 \emph{r}-magnitude and Gaia-DR2 \emph{G}-band magnitudes, as a function of $(X,Y)$ position in the focal plane.
  Stars in the range $18<r<19$ with $0.7<(g-i)_0<1.5$ are used, on the flat part of the relation shown in Fig.~\ref{fig:gaiacalib}. The top row shows the results for the tiles in KiDS-N, the bottom row for those in KiDS-S. {\sc theli} MAG{\_}AUTO is shown on the left, MAG{\_}GAAP{\_}r on the right.}
\label{fig:illumcheck}
\end{figure}

Compared to \citetalias{dejong/etal:2017}, the main change for DR4 is
the inclusion of the photometric illumination correction. This additional processing step is included in the photometric calibration
procedure described in sect.~3.1 (item 4) of
\citetalias{dejong/etal:2017}. After obtaining a photometric
zeropoint, extinction coefficient and colour term of OmegaCAM data
overlapping with SDSS, we measure the residual systematic differences between OmegaCAM
and SDSS-magnitudes over the OmegaCAM field-of-view. These differences
are fitted and corrected with a second-order, two-dimensional polynomial over the
OmegaCAM field-of-view -- see also Fig.~\ref{fig:illum_theli}. The
correction, which like the other calibration images is determined separately for each two-week `observing run' that is processed by {\sc theli}, is directly applied to the single-frame pixel-data. Figure~\ref{fig:illumcheck} compares stellar Gaia \emph{G} magnitudes and the \emph{r}-band magnitudes from the {\sc theli} and {\sc Astro-WISE} reductions, as a function of position on the focal plane. The small residuals on the order of 0.02 mag are caused by different ways of treating the region of the images that are affected by scattered light shadows from the bond wire baffles above the CCD mosaic (see \citetalias{dejong/etal:2017} for more details).

Another, more minor, change to the {\sc theli} workflow is the streamlining of the procedure for masking residual satellite trails on the single exposures. Whereas previously this was done per CCD, the new procedure allows the inspector to mask the track on the entire mosaic in one step.

\section{Single-band \emph{u}, \emph{g}, \emph{r} and \emph{i} catalogues, images and masks}
\label{sec:singlebandcats}

For every co-added image from the {\sc Astro-WISE} pipeline a single-band catalogue was produced using {\sc SExtractor} \citep{bertin/arnouts:1996}, using the same settings as in \citetalias{dejong/etal:2017}\footnote{Note that the {\sc SExtractor} settings are optimised for small sources; measurements for large objects such as extended galaxies should be used with care because of possible shredding or oversubtraction of the background (e.g., \citealt{kelvin/etal:2018}). }.
These catalogues are calibrated photometrically using the nightly zeropoints. To convert the fluxes in these catalogues into SLR+Gaia calibrated magnitudes, the zeropoint given in the DMAG keyword should be used. This value should be added to any magnitude found in the catalogue, and any flux in the catalogue can be turned into a magnitude via
\begin{equation}
m={\rm DMAG}-2.5 \log_{10} {\rm FLUX} .
\end{equation}
The {\sc SExtractor} Kron-like MAG{\_}AUTO and isophotal magnitude MAG{\_}ISO are provided, as well as a range of circular-aperture fluxes, star-galaxy classification and shape parameters. App.~\ref{app:singlebandcat} gives a full list of the parameters included in the catalogues.
Note that the single-band catalogues are derived independently from each co-added KiDS observation, without cross-calibration or source matching across filter bands. In particular, the sources in the \emph{r}-band catalogues are extracted from the {\sc Astro-WISE} co-added images, and differ from those in the nine-band catalogue presented below, which are extracted from the {\sc theli} \emph{r}-band images. Also, sources in the overlap region between adjacent tiles will appear in multiple single-band catalogues, with independent measurements for position, flux, etc. 
The single-band \emph{u}, \emph{g}, \emph{r} and \emph{i} catalogues contain an average of 22k, 79k, 125k, and 65k sources, respectively, per KiDS tile.

DR4 includes the co-added images with corresponding weight maps and masks, as well as the single-band catalogues described above with {\sc SExtractor} output. The {\sc pulecenella} masks identifying stellar reflection haloes are generated in the same way as for \citetalias{dejong/etal:2017}, and are described there.

Though it is possible to generate spectral energy distributions by matching the sources in the \emph{ugri} catalogues for any given tile, it should be remembered that the PSF differs across the bands, and from tile to tile. A better way of obtaining reliable colours is to use the matched-seeing, matched-aperture catalogues described in the next section.

\section{The joint KiDS-VIKING nine-band catalogue}
\label{sec:ninebandcat}
Besides more than doubling the area of sky covered with respect to DR3, the major new feature of KiDS-ESO-DR4 is the inclusion of near-infrared fluxes, from the VIKING survey \citep{edge/etal:2013}. This survey was conceived together with KiDS, as a means to improve knowledge of the spectral energy distributions (SEDs) of the sources, and in particular to enhance the quality of the photometric redshift estimates. Because VISTA entered operations before the VST, VIKING was completed first: full coverage of the $1350\deg^2$ area was reached in August of 2016, while repeat observations of low-quality data were completed in February of 2018\footnote{The VIKING repeat observations taken by September 26, 2016 are incorporated into the DR4 catalogues presented here.}. DR4 includes 
photometry for all sources detected on the \emph{r}-band co-added images as processed with the {\sc theli} pipeline:  \emph{r}-band Kron-like MAG{\_}AUTO, isophotal MAG{\_}ISO magnitudes and a range of circular-aperture fluxes, as well as nine-band optical/near-infrared {\sc GAaP} fluxes for SED estimation. Note that only \emph{r}-band total magnitudes are supplied.

Full details of the near-IR photometry are presented in \cite{wright/etal:prep}. Briefly summarized, the measurements start from the `pawprint' images processed by the Cambridge Astronomical Survey Unit (CASU), which combine each set of jittered observations (offsets of a few arcseconds) into an astrometrically and photometrically calibrated set of sixteen detector-sized images. Each VIKING tile consists of six such pawprints, with observations offset by nearly a full detector width in right ascension, and half that in declination. Because the large gaps between the detectors can result in significant PSF quality jumps after co-addition, the VIKING photometry for KiDS-ESO-DR4 is performed by running the PSF Gaussianization and {\sc GAaP} separately on each pawprint detector. The final flux in each VISTA band is the optimally weighted average of the individual flux measurements, using the individual flux errors that are derived by propagating the pixel errors (including covariance) through the {\sc GAaP} procedure as descibed in appendix~A of \cite{kuijken/etal:2015}. Due to the VIKING observing strategy sources typically appear on two VIKING pawprints (four in case of \emph{J}-band), but a few percent of the sources appear six times (twelve in \emph{J}) or more. The number of exposures that contribute to each source's VIKING fluxes is given in the catalogue.
Note that the optical photometry is performed on the co-added images, which are much less sensitive to PSF changes between sub-exposures because of the very high pixel coverage fraction of the focal plane of the optical camera (which has minimal gaps between the three-edge buttable CCDs in the instrument).

Because of their respective cameras' different footprints on the sky, KiDS and VIKING tile the sky differently. Data quality variations (depth, seeing, background...) follow a square-degree pattern for the KiDS data, and a $1.5\times1$ degree pattern for VIKING; moreover within VIKING tiles the variations can be more complex because of the larger gaps between VIRCAM detectors.

\begin{figure}
    \includegraphics[width=0.5\textwidth]{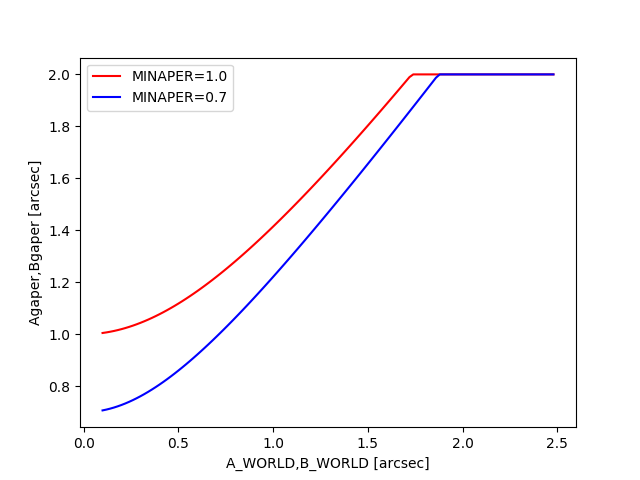}
  \caption{Relation between the {\sc SExtractor} major and minor axis measurements ($A$ and $B$) and the adopted {\sc GAaP} apertures, for the two minimum aperture (MIN{\_}APER) values that have been used.}
  \label{fig:minaperformula}
\end{figure}

The essence of the {\sc GAaP} photometry \citep{kuijken/etal:2015} contained in the DR4 nine-band catalogue is to provide accurately aperture-matched fluxes across all wavebands, properly corrected for PSF differences. The aperture major and minor axis lengths, Agaper and Bgaper, are set from the {\sc SExtractor} size and shape parameters measured on the detection image, via
\begin{equation}
  X\hbox{gaper}=\left({X\rm\_WORLD}^2+{\rm MIN\_APER}^2\right)^{1/2}\qquad\hbox{for}\quad X=A,B
\end{equation}
(see Fig.~\ref{fig:minaperformula}) with the position angle equal to the position angle THETA{\_}WORLD from the detection image.\footnote{The convention in the {\sc GAaP} code is that the position angle is measured from East to North, so the catalogue contains the angle PA{\_}GAAP$=180-$THETA{\_}WORLD.} As it was in \citetalias{dejong/etal:2017}, the MIN{\_}APER parameter is set to $0\farcs7$ for all sources, and in addition $A$gaper and $B$gaper are maximized at 2\arcsec.
Imposing a maximum aperture size helps to ensure that the {\sc GAaP} colours are not contaminated by neighbouring sources, but we do not attempt here to deblend overlapping sources. We note that an explicit flagging of sources affected by neighbours is done by the {\sc SExtractor} source detection step, and is also an important part of the forthcoming {\em lens}fit shape measurements.
For further discussion of the choice of {\sc GAaP} aperture size, see \cite{kuijken/etal:2015}, appendix~A2.

\begin{figure}
\includegraphics[width=0.5\textwidth]{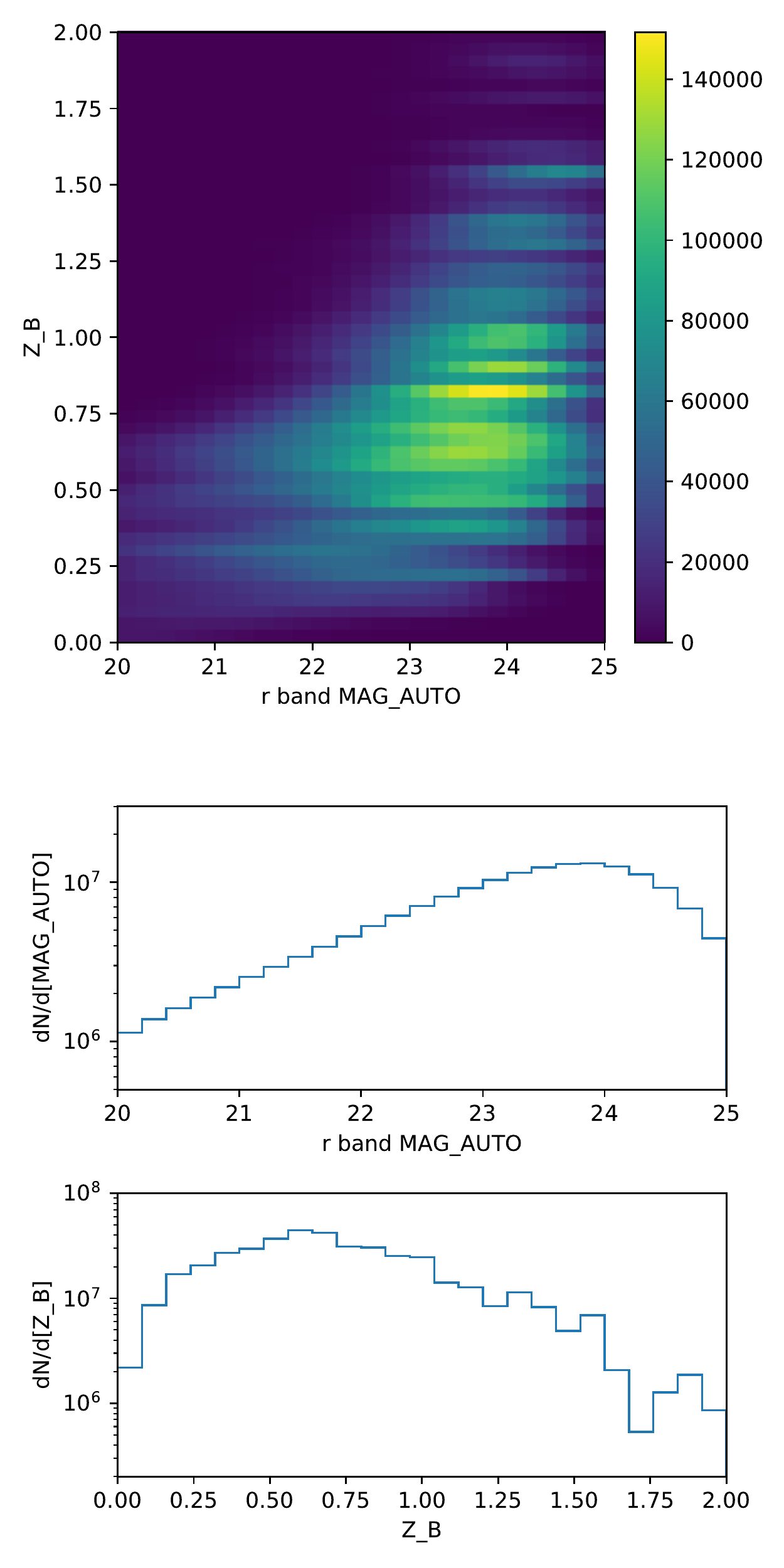}
\caption{Joint and marginal histograms of the $r$-band MAG\_AUTO magnitude and the Z\_B photometric redshift estimate for the sources in the nine-band catalogue.}
\label{fig:z-mag-dist}
\end{figure}

\begin{figure*}
  \includegraphics[width=\textwidth,bb=0 2.4in 12in 12in]{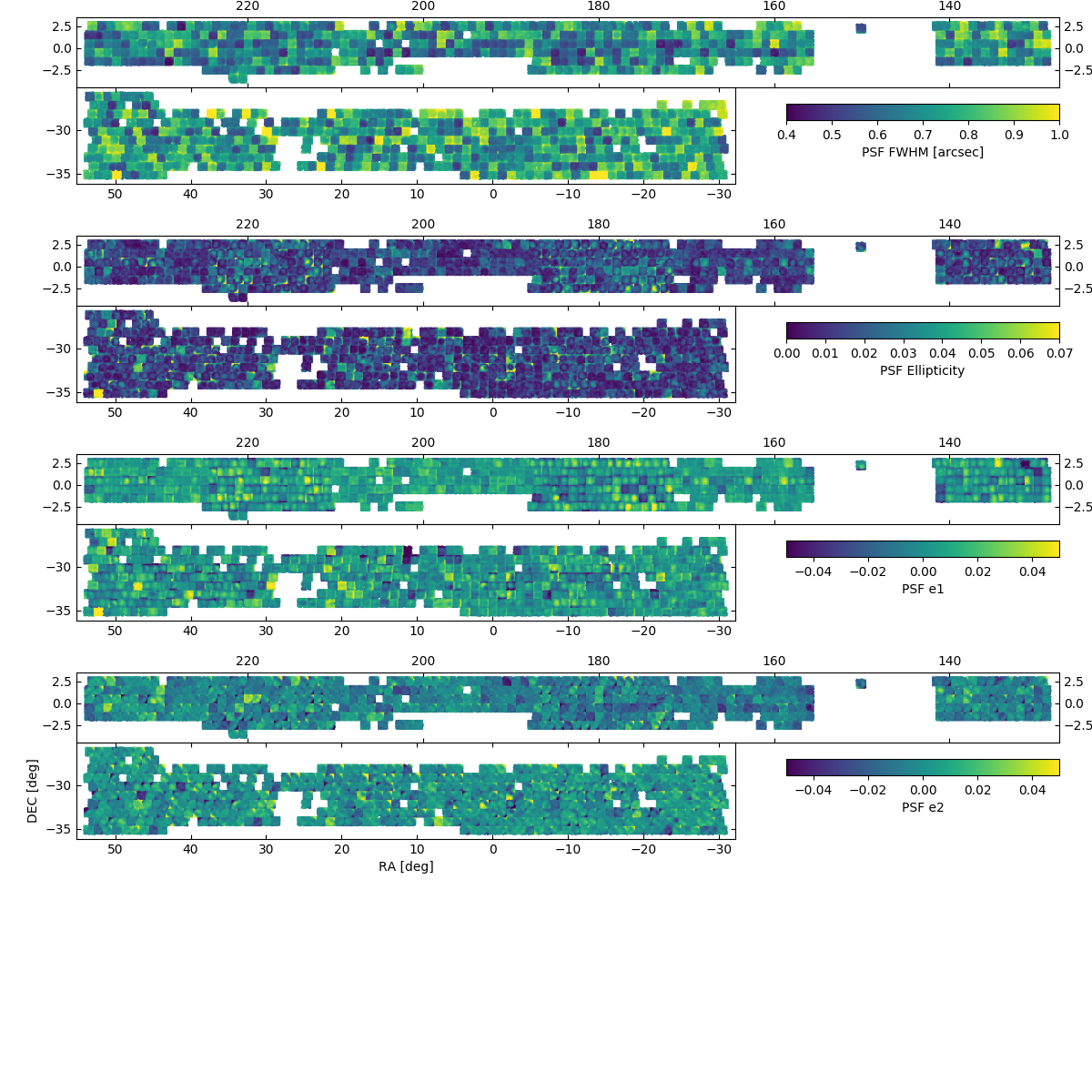}
  \caption{\emph{r}-band PSF properties across KiDS-ESO-DR4. From top to bottom: FWHM, ellipticity modulus, and ellipticity components 1 and 2 (elongation along the pixel $x$ axis and diagonal, respectively).}
  \label{fig:psfmaps}
\end{figure*}

In rare cases, no {\sc GAaP} flux can be measured with this setup, because
{\sc GAaP} photometry can only be determined when the specified aperture size is larger than the Gaussianized PSF. To provide colours for these sources, DR4 includes a second run with larger apertures, obtained by setting MIN{\_}APER=1\farcs0. The results from both {\sc GAaP} runs are present in the catalogues, with keywords whose names contain {\_}0p7 and {\_}1p0 respectively. In general the 1p0 fluxes will have a larger error than those from the standard 0p7 setup, since the larger aperture includes more background noise. However, when the Gaussianized PSF size is close to that of the {\sc GAaP} aperture, the error increases\footnote{because the aperture that is used for the photometry is the {\sc GAaP} aperture deconvolved by the PSF}; in such cases it may happen that the larger MIN{\_}APER=1\farcs0 leads to a smaller flux error (and when the PSF is too broad the flux error is formally infinite). In the DR4 catalogue a source-by-source decision is made which optimal MIN{\_}APER choice to use as input for the photometric redshifts, as follows:
\begin{enumerate}
\item For all bands x, calculate the flux error ratios
\begin{equation}
R_{\rm x} = {\rm FLUXERR\_GAAP{\_}1p0{\_}x} / {\rm FLUXERR{\_}GAAP{\_}0p7{\_}x}
\end{equation}
\item If
\begin{equation}
\max_{\rm x} (R_{\rm x}) \times \min_{\rm x} (R_{\rm x}) < 1
\end{equation}
then use the 1p0 fluxes for this source, else adopt 0p7.
\end{enumerate}
This choice ensures that the smaller 0p7 aperture is used, unless there is a band for which the larger aperture gives a smaller flux error, for the reasons indicated above: in that case, the 1p0 fluxes are used if the fractional reduction in the error in that band is greater than the fractional penalty suffered by the other bands.\footnote{This formulation also handles the convention that an unmeasured {\sc GAaP} flux returns an error of $-1$, provided the rare cases that a 1p0 flux cannot be measured but a 0p7 flux can, are caught.} The larger MIN{\_}APER is preferred in some four percent of the cases.
These optimal {\sc GAaP} fluxes are reported in the catalogue with the FLUX{\_}GAAP{\_}x keywords.

The flux and magnitude zeropoints in the catalogue are as follows.
\begin{itemize}
\item All {\sc GAaP} fluxes are reported in the units of the image pixel ADU values. For the KiDS images these correspond approximately to a photometric AB magnitude zeropoint of 0; for the VISTA data this zeropoint is 30.
\item The {\sc GAaP} magnitudes MAG{\_}GAAP{\_}0p7{\_}x and MAG{\_}GAAP{\_}1p0{\_}x for the KiDS \emph{ugri} bands are calculated from the corresponding fluxes using the zeropoint DMAG{\_}x or DMAG{\_}x{\_}1 (see Sect.~\ref{sec:photcal}) and recorded in the catalogue header.
\item The {\sc GAaP} magnitudes for the VIKING \emph{ZYJHK}$_{\rm s}$ bands are calculated with the zeropoint 30.
\item The optimal {\sc GAaP} fluxes FLUX{\_}GAAP{\_}x are equal to one of the 0p7 or 1p0 sets, as described above.
\item The optimal {\sc GAaP} magnitudes MAG{\_}GAAP{\_}x are calculated as above, but in addition are also corrected for Galactic extinction by subtracting the EXTINCTION{\_}x value (obtained using the data in Table~\ref{tab:extinction}).
\item The colours COLOUR{\_}GAAP{\_}x{\_}y in the catalogue are obtained as differences between these extinction-corrected MAG{\_}GAAP{\_}x magnitudes, and are therefore corrected for Galactic reddening.
\end{itemize}

The nine-band catalogue also contains those KiDS sources that fall outside the VIKING footprint, as can be seen on the maps of limiting magnitude in Sect.~\ref{sec:depth} below. 
Note in particular that the COSMOS field near (RA,DEC)=$(150\fdg0,2\fdg5)$ (KiDS-N-D2 in Table~\ref{tab:fields}) is not part of the VIKING survey, though other -- in some cases much deeper -- VISTA data do exist for this field. These near-infrared data do not form part of KiDS-DR4, but they are incorporated in the calibration of the photometric redshifts for the KV450 analysis in \cite{hildebrandt/etal:prep}.

\begin{figure}
  \includegraphics[width=0.5\textwidth]{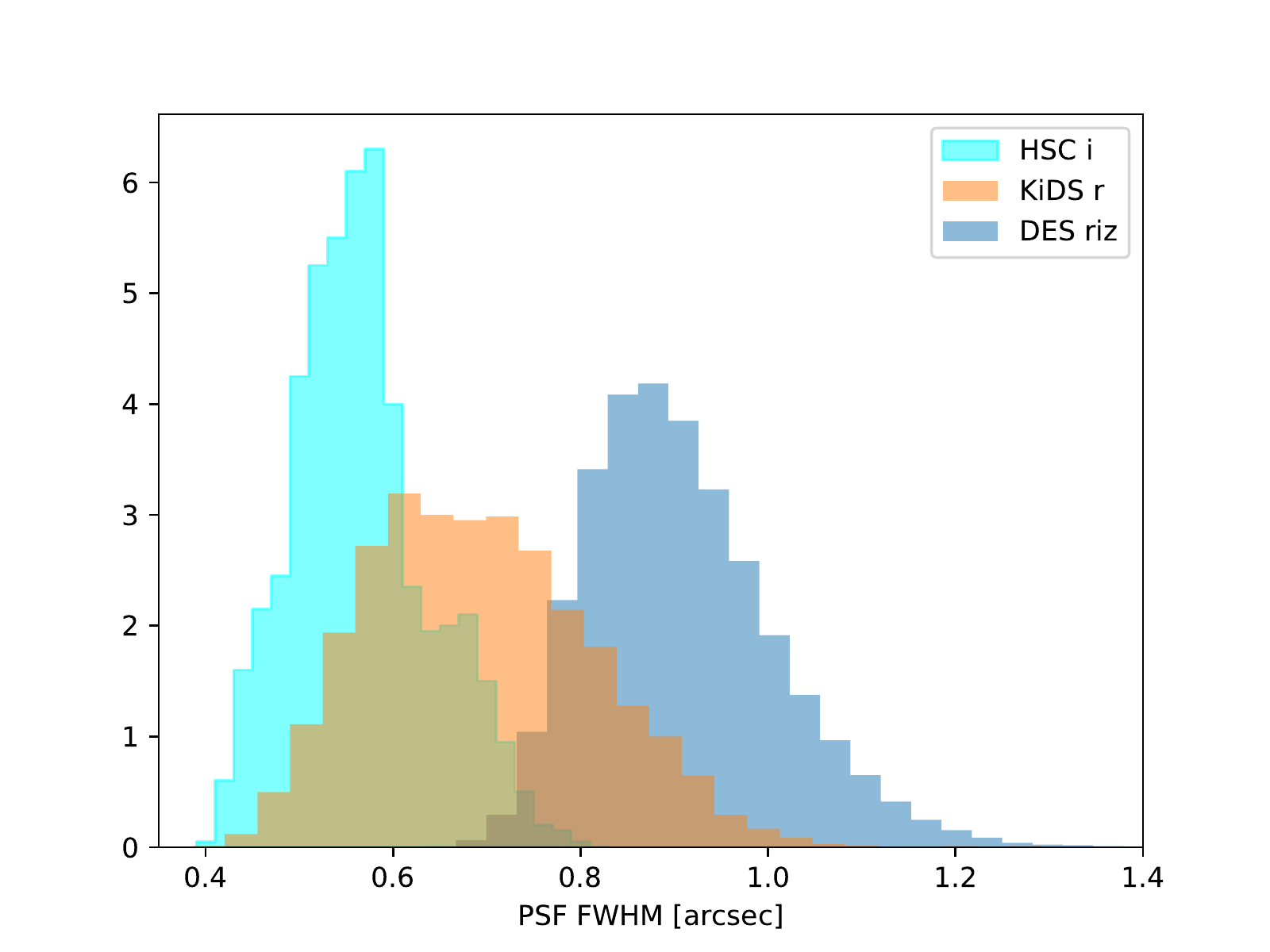}
  \caption{The KiDS \emph{r}-band seeing FWHM distribution (red) compared to those used for shape measurements in the other major ongoing weak lensing surveys: HSC \emph{i} band (cyan) and DES \emph{r}+\emph{i}+\emph{z} (blue).}
  \label{fig:fwhmDESKiDSHSC}
\end{figure}

Fig.~\ref{fig:z-mag-dist} illustrates the $r$-band depth and photometric redshift distribution of the sources in the nine-band catalogue.
For a list of all columns, see App.~\ref{app:ninebandcat}. The catalogue contains just over 100 million objects.

\section{Data quality}
\label{sec:quality}

\subsection{Image quality}
For weak lensing, the most critical science case for KiDS, image quality is a crucial property of the data. KiDS scheduling is designed to take advantage of the periods of excellent seeing on Paranal, by prioritising the \emph{r}-band exposures at those times. The resulting seeing distribution of the four KiDS bands was shown in Fig.~\ref{fig:dataquality}.

In Fig.~\ref{fig:psfmaps} we present maps of the KiDS-ESO-DR4 \emph{r}-band PSF size and ellipticity, obtained by running the {\em lens}fit PSF modelling code used for the KiDS-450 analysis \citetalias{hildebrandt/etal:2017} on the {\sc theli} images. The top row of the Figure shows the PSF FWHM. Individual KiDS tiles are clearly visible. The second row shows the PSF ellipticity, and the bottom two rows the `1' and `2' ellipticity components. The results of the VST improvements that were implemented in 2015 are reflected in the DR4 data: PSF variations are significantly reduced in the newly added data compared to the data from DR1+2+3 (see the blue areas in Fig.~\ref{fig:fields}).

 In Fig.~\ref{fig:fwhmDESKiDSHSC} we compare the seeing distribution of the KiDS-DR4 \emph{r}-band data with those of the images used for the lensing measurements in the other major ongoing surveys: the DES-Year 1 \emph{riz} data \citep{zuntz/etal:2018}, which has a similar depth to KiDS, and the HSC DR1 \emph{i}-band lensing catalogue \citep{mandelbaum/etal:2018}. The superior seeing of KiDS compared to DES explains why both surveys are providing cosmic shear constraints of comparable power, despite the larger DES area. The impact that can be expected from the complete HSC survey is also evident, even more so considering that it is significantly deeper than KiDS and DES.

\subsection{Astrometry}

\begin{figure}
  \includegraphics[width=0.5\textwidth]{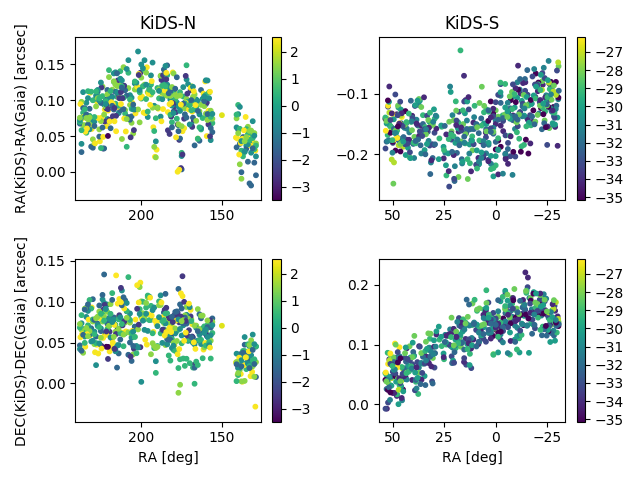}
  \caption{Median {\sc theli} astrometry residuals per KiDS tile in the nine-band catalogues, as measured from unsaturated Gaia stars, in arcseconds. Left/right plots show KiDS-N/S, and the top and bottom rows show the median offsets per tile in RA and DEC. The colours indicate the declination of the tiles, in degrees.}
  \label{fig:astrometryTheliGaia}
\end{figure}
\begin{figure}
  \includegraphics[width=0.5\textwidth]{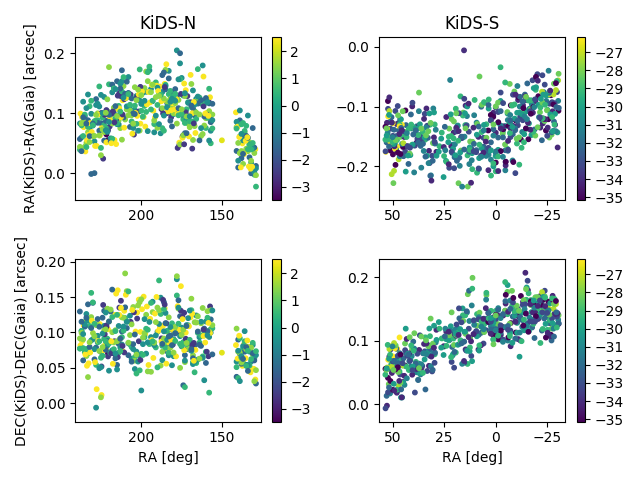}
  \caption{As Fig.~\ref{fig:astrometryTheliGaia}, but for the {\sc Astro-WISE} single-band \emph{r} catalogues.}
  \label{fig:astrometryAWrbandGaia}
\end{figure}

The astrometry in the {\sc theli}-processed \emph{r}-band detection images and catalogues is tied to SDSS in the North, and to 2MASS in the South. The {\sc Astro-WISE} images and single-band catalogues are all tied to 2MASS, as is VIKING. Slight differences exist between these two reference catalogues.

We compare the {\sc theli} and {\sc Astro-WISE} astrometry to the Gaia DR2 data in Figs.~\ref{fig:astrometryTheliGaia} and \ref{fig:astrometryAWrbandGaia}. Systematic residuals between either catalogue and Gaia are at the level of 200 mas; between the {\sc theli} and {\sc Astro-WISE} reductions the differences are at most 50 mas. These latter differences are sufficiently small that they will not affect the {\sc GAaP} photometry (where the apertures are defined on the {\sc theli} images, but the fluxes measured at the corresponding positions in the {\sc Astro-WISE} images).

\begin{figure}
\includegraphics[width=0.5\textwidth]{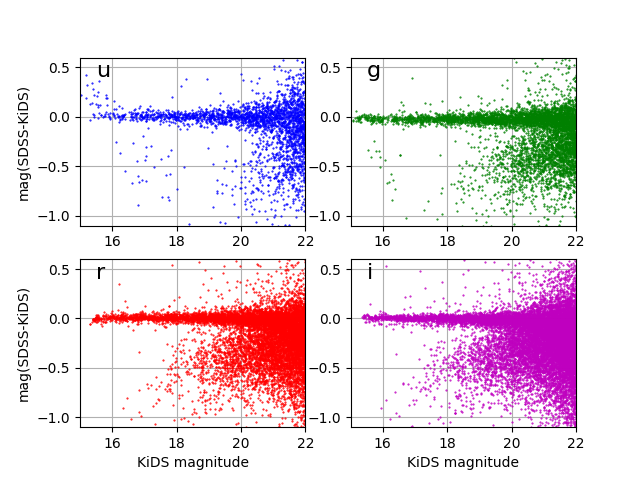}
\caption{Comparison between the SDSS DR9 model magnitudes and KiDS-ESO-DR4 {\sc GAaP} photometry. The comparisons are shown for \emph{u}, \emph{g}, \emph{r}, and \emph{i} bands, for an example tile (KIDS{\_}188.0{\_}$-$0.5).}
\label{fig:photvsdss-tile}
\end{figure}

\begin{figure*}
\includegraphics[width=\textwidth]{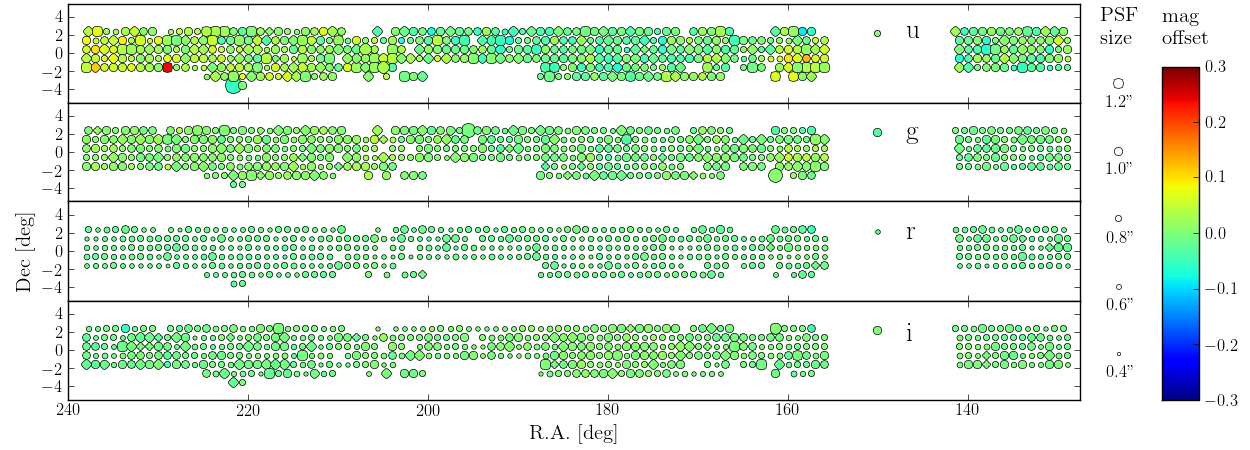}
\caption{Photometric comparison between KiDS and SDSS photometry of stars in the KiDS-N area. The colourscale indicates the mean magnitude offset $m_{\rm KiDS}-m_{\rm SDSS}$ of high-signal stars $u<20$, $g<22$, $r<22$, $i<20$ in every tile. The size of the symbol increases with the FWHM of the PSF. The one outlier in the \emph{u}-band photometry, tile KIDS{\_}229.0{\_}-2.5{\_}u, is one of the highest-extinction fields in the survey and contains few \emph{u}-band stars.}
\label{fig:KiDSN-SDSS-photom}
\end{figure*}

\subsection{Photometry}
\label{sec:photometry}

We assess the accuracy of the KiDS photometric calibration by comparing with overlapping, shallower surveys. For galaxies, comparing the KiDS photometry with catalogues from other surveys is complex, since for extended sources the {\sc GAaP} fluxes do not measure total fluxes (see Sect.~\ref{sec:gaap}). If desired, the \emph{r}-band circular aperture fluxes in the catalogue can be used to generate a curve-of growth, and total magnitudes can then be estimated from the {\sc GAaP} colours. (Alternatively, the \emph{r}-band MAG{\_}AUTO can be combined with the {\sc GAaP} colours to generate estimated Kron-like magnitudes in the other bands.
Such procedures assume that there are no colour gradients in the galaxy, an assumption that could be tested by comparing the 0p7 and 1p0 {\sc GAaP} fluxes.)

For stars and other unresolved objects, the situation is different, since their {\sc GAaP} fluxes {\em are} total fluxes: the {\sc GAaP} flux with Gaussian aperture function $W=\exp[-\frac12(x^2/A^2+y^2/B^2)]$ (in coordinates centered on the source and rotated to align with the aperture major and minor axes) is the integral (on the pre-seeing sky)
\begin{equation}
  \int {\rm d} x\, {\rm d} y\, I(x,y) \, W(x,y)
\end{equation}
which evaluates to the flux $F$ of the source when the intensity $I(x,y)$ is $F$ times a delta function.

Such a comparison is shown in Fig.~\ref{fig:photvsdss-tile}, for an example tile in KiDS-N where SDSS and KiDS overlap. As expected, the stars form tight sequences close to the line of zero magnitude difference, demonstrating that the KiDS and SDSS zero points are consistent for this tile, while the KiDS {\sc GAaP} magnitudes of galaxies trail towards fainter magnitudes than the corresponding SDSS magnitude. A similar comparison for the near-IR data is presented in \cite{wright/etal:prep}.
Tile-by-tile consistency in the four bands, for the KiDS-SDSS overlap, is shown in Fig.~\ref{fig:KiDSN-SDSS-photom}. The larger scatter in the \emph{u} band, already discussed in Sect.~\ref{sec:photcal}, is evident, particularly in the fields with higher extinction at the extremes of the RA range. In these tiles  there are fewer stars with reliable \emph{u} band photometry; in addition, because of their lower Galactic latitude the foreground extinction screen approximation is less well justified.

\begin{figure}
  \includegraphics[width=0.5\textwidth]{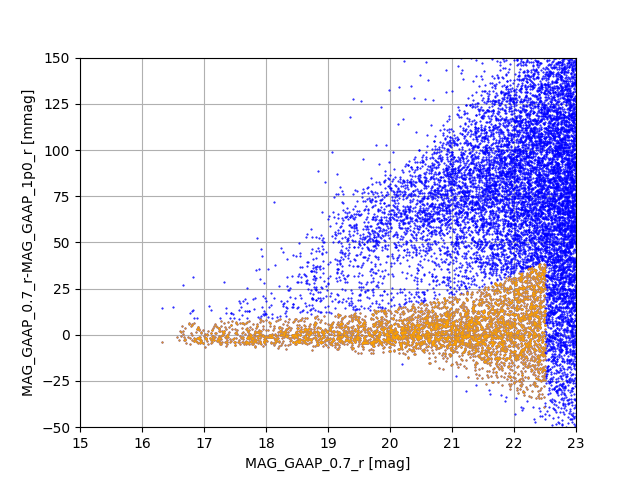}
  \includegraphics[width=0.5\textwidth]{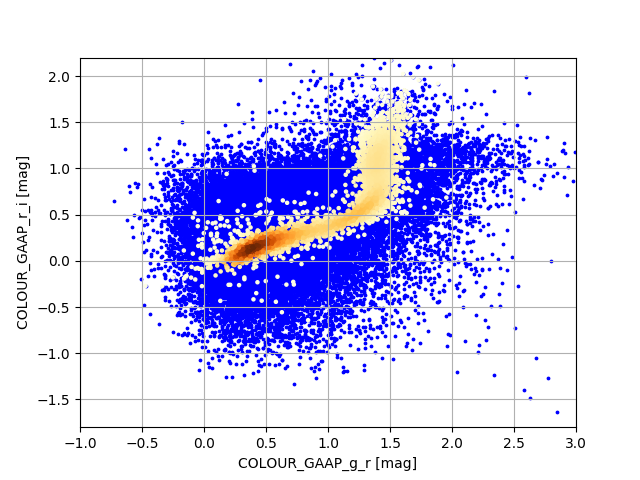}
    \caption{Top: star-galaxy separation using the 0p7 and 1p0 {\sc GAaP} magnitudes. The objects shown in gold form a sequence along which both apertures yield consistent fluxes, indicating that they are unresolved. The bottom plot shows the $g-r,r-i$ colour-colour diagram for the same sources, confirming that these sources are stars.}
  \label{fig:gaapstargal}
\end{figure}

Figure \ref{fig:gaapstargal} illustrates an internal consistency check of the {\sc GAaP} magnitudes, that can also serve as a new star-galaxy classifier. The top panel shows the difference between the `0p7' and `1p0' \emph{r}-band {\sc GAaP} magnitudes in an example tile. This comparison clearly reveals two populations which are well-separated at the bright end, down to magnitude 22.5. A conservative cut, shown in gold, identifies unresolved objects, for which the {\sc GAaP} magnitude is independent of aperture size. (Most of the remaining objects are resolved, with  higher fluxes for the larger aperture.) The bottom panel shows the location in the $g-r,r-i$ colour-colour diagram of the two populations, confirming that the unresolved objects have mostly stellar colours, whereas the others show the colour distribution expected of a population of galaxies at a wide range of redshifts.

\begin{table}
  \caption{Settings for the BPZ photometric redshift calculations}
  \label{tab:bpz}
  \centering
  \begin{tabular}{lr}
    \hline\hline
    Parameter & Value\\
    \hline
    BPZ version &  1.99.3\\
    ZMAX & 7.0 \\
    INTERP & 10 \\
    ODDS & 0.68 \\
    MIN{\_}RMS & 0.067\\ 
    PHOTO\_ERRORS & yes\\
    \hline
  \end{tabular}\\
  \end{table}

\subsection{Photometric redshifts}
\label{sec:photoz}
The nine-band catalogue contains photometric redshift estimates, obtained with the BPZ code \citep{benitez:2000}. It gives the most probable redshift values, as well as the `$1\sigma$' 32nd and 68th percentiles of the posterior probability distributions, and the best-fit SED type. The BPZ version and settings are given in Table~\ref{tab:bpz}.

\begin{figure}
  \includegraphics[width=0.5\textwidth,bb=-25 0 325 525]{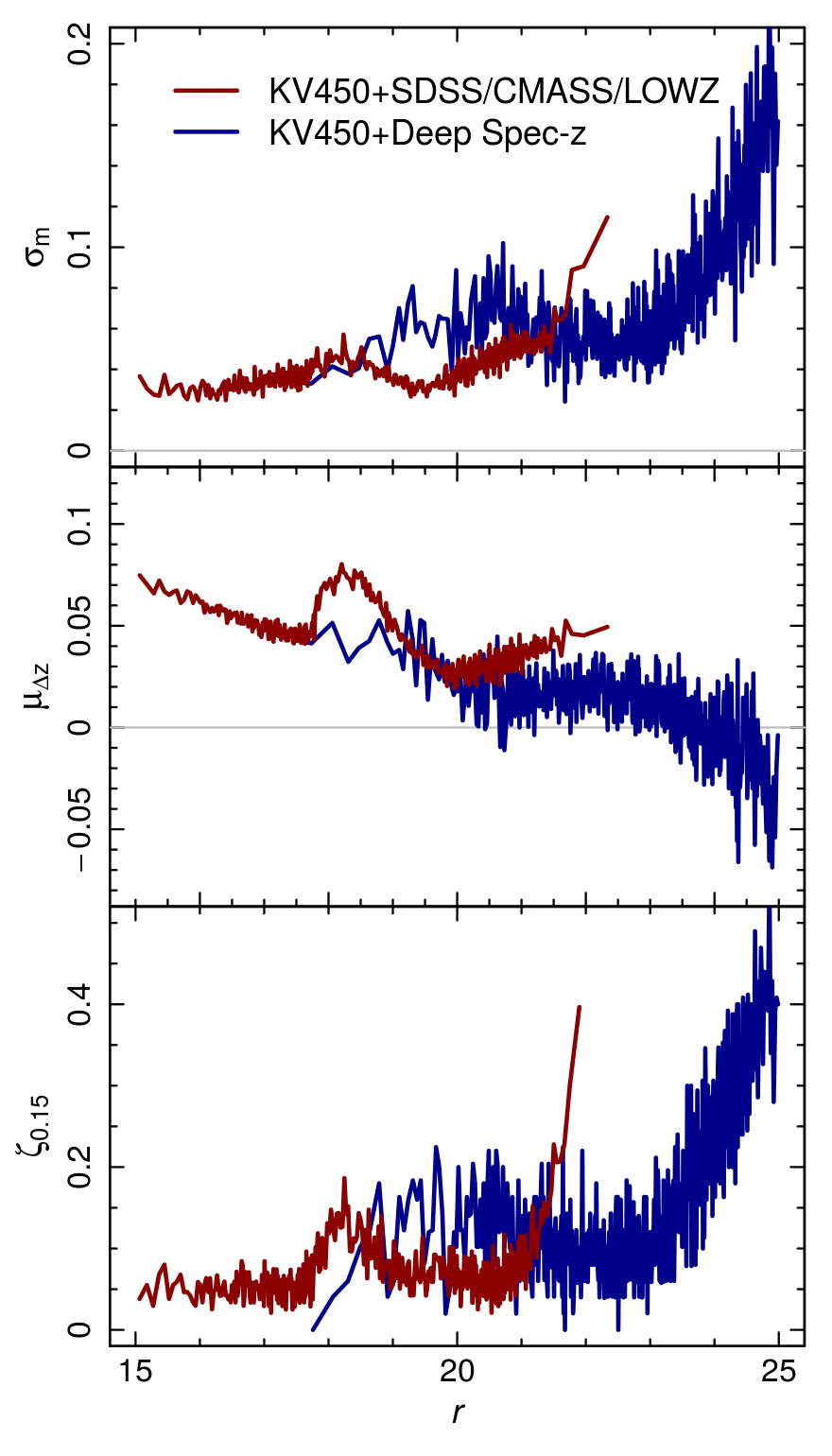}
  \caption{\emph{r}-Magnitude dependence of KiDS DR4 photo-$z$ statistics based on the findings of \cite{wright/etal:prep} for the deep spec-$z$ fields (blue lines) and a direct comparison of DR4 photo-$z$ and SDSS/2dFLenS spec-$z$ (red lines). The top panel shows the normalised-median-absolute-deviation of the quantity $\Delta_z/(1+z)$, the middle panel shows the mean $\mu_{\Delta z}$ of that quantity, and the lower panel shows the rate $\zeta_{0.15}$ of outliers with $|\Delta_z/(1+z)|\ge 0.15$. (Note that this definition of $\zeta_{0.15}$ exaggerates the outlier fraction when $\sigma_{\mathrm m}$ approaches 0.1.) The scatter between neighbouring points gives an indication of the error bars on these quantities.}
  \label{fig:zz_stats}
\end{figure}

Since \citetalias{dejong/etal:2017} we have implemented several changes to our photo-$z$ setup. We updated the prior redshift probability used in BPZ to the one given in \cite{raichoor/etal:2014}. This prior reduced uncertainties and catastrophic failures for faint galaxies at higher redshifts, but appears to generate a redshift bias for bright, low-redshift galaxies. We therefore caution users of the catalogue to calibrate the BPZ redshifts appropriately before using them. At bright magnitudes, where complete training data are available, it is advantageous to use an empirical photo-$z$ technique like the one presented for the \citetalias{dejong/etal:2017} data set in \citet{bilicki/etal:2018}; 
specific selection and calibration of luminous red galaxies \citep[LRG,][]{vakili/etal:prep} is also effective. Bright and LRG samples based on DR4 and taking advantage of its unique, deep, nine-band coverage are in preparation.

Another change is related to the photo-$z$ errors and the ODDS quality indicator. In previous releases we reported 95\% confidence intervals for the Bayesian redshift estimate. With DR4 we switch to 68\% confidence intervals as mentioned above, which requires some changes to the settings and also changes the values of the ODDS parameter. The changes in prior and BPZ settings, together with the fact that we are using full nine-band photometry in a KiDS data release for the first time mean that previous photo-$z$ results based on optical-only photometry \citep[e.g.,][]{kuijken/etal:2015} are no longer advocated.

\begin{figure}
  \includegraphics[width=0.5\textwidth]{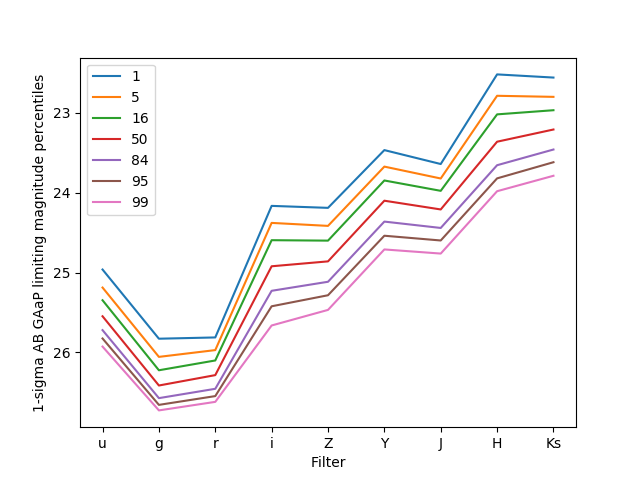}
  \caption{Various percentiles of the 1-$\sigma$ {\sc GAaP} limiting magnitudes for the nine wavelength bands. The width of the distributions is driven by differences in seeing, air mass and sky brightness across the KiDS and VIKING surveys.}
  \label{fig:limmag}
\end{figure}

Further discussion of the nine-band KiDS+VIKING photometric redshifts, as well as a comparison to the \citetalias{hildebrandt/etal:2017} photo-$z$, is provided in \cite{wright/etal:prep} where a similar setup\footnote{
The main difference of the DR4 setup is the implementation of two different minimum aperture sizes as discussed in Sect.~\ref{sec:ninebandcat}. As this change only impacts data with seeing that greatly varies between bands in \cite{wright/etal:prep}, it does not affect the photo-$z$ statistics presented here.} 
was used. There the BPZ photo-$z$ are tested against several deep spectroscopic surveys. At full depth ($r\la24.5$) the photo-$z$ show a scatter (normalised-median-absolute-deviation) of $\sigma_{\rm m}=0.072$ of the quantity $\Delta_z/(1+z)=(z_{\rm B}-z_{\rm spec})/(1+z_{\rm spec})$ and a fraction $\zeta_{0.15}=17.7\%$ of outliers with $|\Delta_z/(1+z)|\ge 0.15$. The magnitude dependence of these quantities is shown in Fig.~\ref{fig:zz_stats}. The effect of the different  selection criteria in the spectroscopic catalogues is evident (e.g., the bump in the redshift bias $\mu_{\Delta z}$ near $r=18$, which marks the transition from the BOSS LOWZ to CMASS samples), illustrating that calibrating the photometric redshift error distribution requires care (see \citealt{bilicki/etal:2018} and the extensive discussion of direct calibration techniques in \citetalias{hildebrandt/etal:2017} for further details). Note also that the photo-z setup was optimised for the fainter, $r>20$ galaxies that are of interest for the KiDS cosmology analysis, and not for the brighter galaxies.

\begin{figure*}
\centering
\includegraphics[width=\textwidth,bb=0 2.4in 12in 12in]{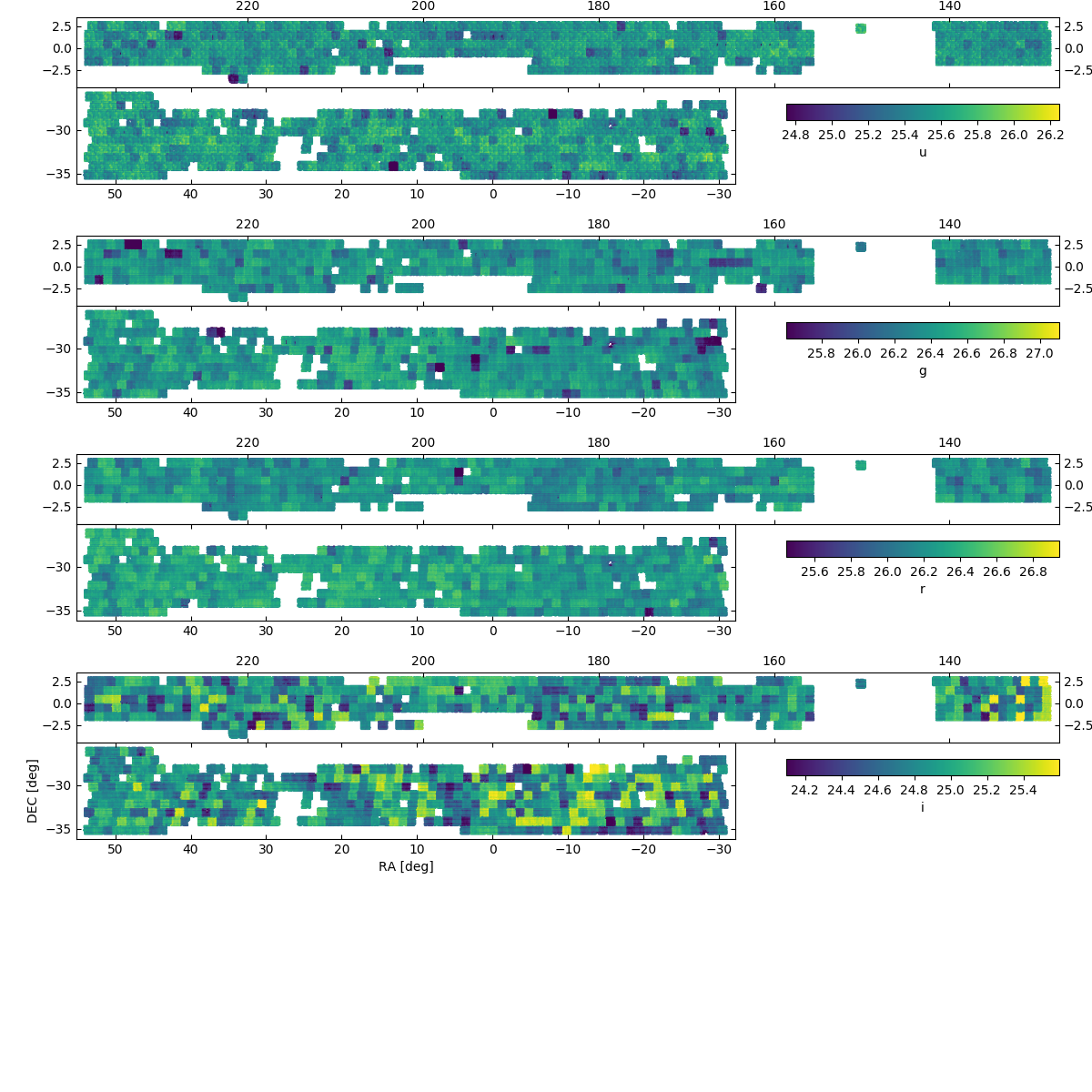}
  \caption{Maps of the median limiting {\sc GAaP} magnitude, corresponding to the 1-$\sigma$ flux error, in $0\fdg1\times0\fdg1$ cells, for the four KiDS filter bands. The colour scale in every map spans $\pm0.75$ magnitude about the median. Note the significantly greater inhomogeneity of the \emph{i}-band data: this is expected to improve in the final data release after a second pass is completed.}
  \label{fig:limmagmapsKiDS}
\end{figure*}
\begin{figure*}
\centering
\includegraphics[width=\textwidth]{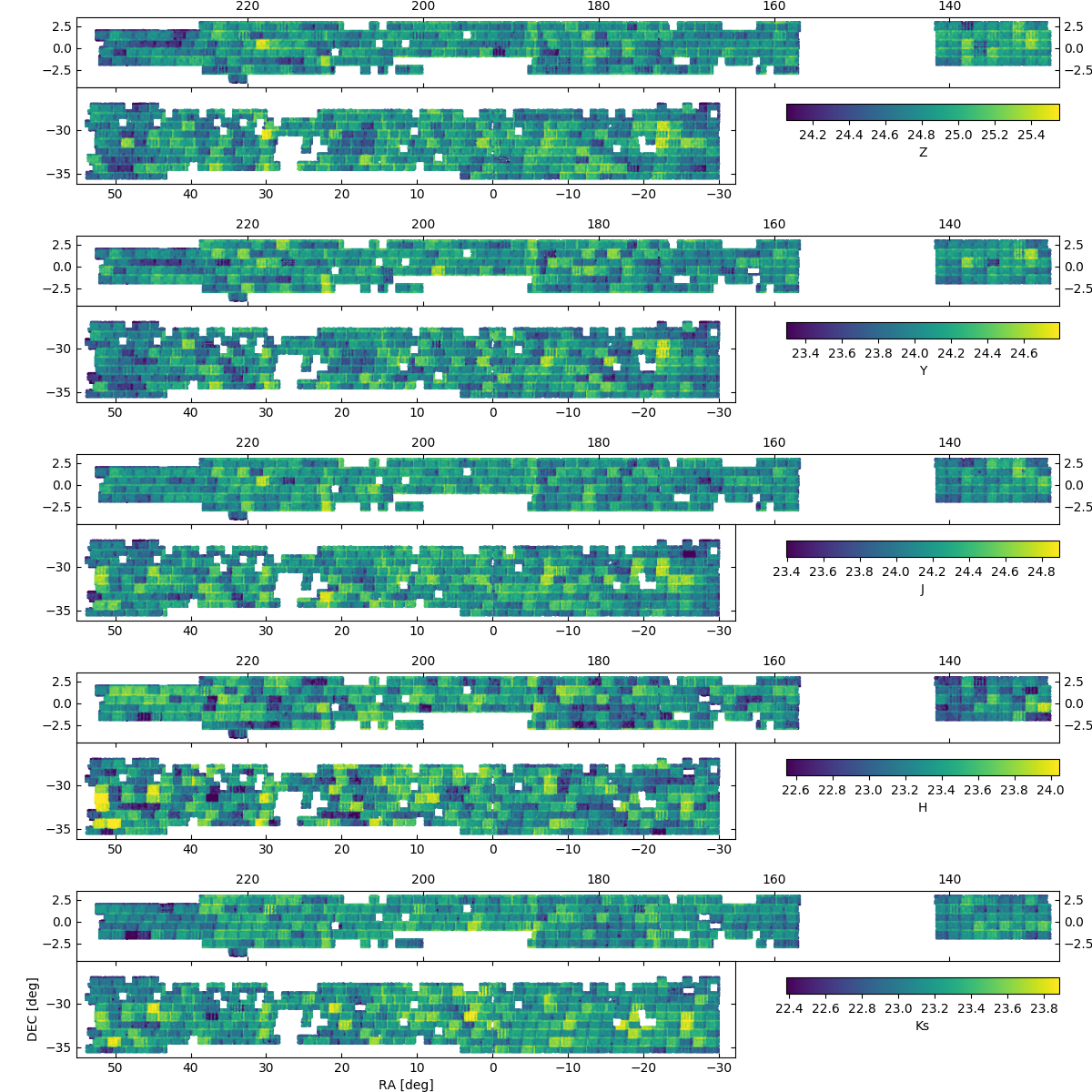}
  \caption{Maps of the \emph{r}-band selected sources'  median limiting {\sc GAaP} magnitude, corresponding to the 1-$\sigma$ flux error, in $0\fdg1\times0\fdg1$ cells, for the five VIKING bands. Note the rectangular $1\fdg5\times1\fdg0$ patterns, due to the footprint of the VIRCAM instrument.}
  \label{fig:limmagmapsVIKING}
\end{figure*}

\subsection{Photometric depth and homogeneity}
\label{sec:depth}
Figure \ref{fig:limmag} shows the distribution of the different bands' 1-$\sigma$ limiting magnitudes in the catalogue. Note the narrow range of the limiting magnitudes in the \emph{u}, \emph{g}, and particularly the \emph{r} band, a consequence of the KiDS observing strategy of choosing which dark-time band to observe in according to the seeing conditions.
Maps of the median limiting magnitude in $0\fdg1\times0\fdg1$ cells are presented in Fig.~\ref{fig:limmagmapsKiDS} (KiDS \emph{ugri}) and~\ref{fig:limmagmapsVIKING} (VIKING \emph{ZYJHK}$_{\rm s}$).

\section{Data access}
\label{sec:access}

There are several ways through which KiDS-ESO-DR4 data
products can be accessed. An overview is presented in this section and
up-to-date information is also available at the KiDS DR4
website.

The data products that constitute the DR4 release (stacked \emph{ugri}
images and their associated weight maps, flag maps, and single-band
source lists for 1006
survey tiles, the \emph{r}-band detection images and their
  weight maps, as well as the multi-band catalog combining KiDS and
VIKING photometry together with flag maps combining mask information
from all filters),
are released via the ESO Science Archive, and
also accessible via {\sc Astro-WISE} and the KiDS website.

\subsection{ESO science archive}

All main release data products are disseminated through the ESO
Science Archive
Facility\footnote{\url{http://archive.eso.org/cms.html}}, which
provides several interfaces and query forms.  All image stacks, weight
maps, flag maps and single-band source lists are provided on a per
tile basis via the `Phase 3 main query form'. This interface
supports queries on several parameters, including position, object
name, filter, observation date, etc. and allows download of the
tile-based data files. Also the multi-band catalog, which is stored in
per-tile data files, is available in this manner.
A more advanced method to query the multi-band catalog is provided by
the `Catalogue Facility query interface', which enables users to
perform queries on any of the catalog columns, for example
facilitating selections based on area, magnitude, photo-z or shape
information. Finally, data can be queried directly from a new
  graphical sky projection interface known as the `Science Portal'. Query results can subsequently be exported to various
(single-file) formats.

\subsection{{\sc Astro-WISE} archive}

Most of the data products can also be retrieved from the {\sc Astro-WISE}
system \citep{begeman/etal:2013}. This data processing and management
system is used for the production of these data products and retains
the full data lineage. For scientists interested in access to various
quality controls, further analysis tools, or reprocessing of data this
access route may be convenient.
The DBviewer web service\footnote{\url{http://dbview.astro-wise.org}}
allows querying for data products and supports file downloads, viewing
of inspection plots, and data lineage browsing. Links with DBviewer
queries to complete sets of data products are compiled on the KiDS DR4
website. Several data products that are not in the ESO archive may be retrieved through this route: most importantly, the PSF-Gaussianized images and the individual CCD sub-exposures after various stages of processing in the {\sc Astro-WISE} pipeline are available here.

\subsection{KiDS DR4 website}

Apart from offering an up-to-date overview of all data access routes,
the KiDS DR4
website\footnote{\url{http://kids.strw.leidenuniv.nl/DR4}} also
provides alternative ways for data retrieval and quality control.

The synoptic table presents for each observation (tile/filter) a
combination of inspection plots relating to the image and source
extraction quality, as well as links for direct downloads of the
various data FITS files. Furthermore, direct batch downloads of all
DR4 FITS files are supported by supplying \textsc{wget} scripts.

\section{Summary and outlook}
\label{sec:summary}

With the KiDS-ESO-DR4 data release, data for over 1000 square degrees, more than two thirds of the target KiDS footprint, is now publicly available. Co-added images with associated weights and masks, as well as single-band source catalogues, may now be accessed through the ESO archive or the KiDS project website.

Moreover, through a combined analysis of these KiDS images with data from the VIKING survey, a  nine-band matched-aperture \emph{u}--\emph{K}$_{\rm s}$ catalogue containing some 100 million galaxies has been created, with limiting 5-$\sigma$ AB magnitudes ranging from ca.\ 25 in \emph{g} and \emph{r} bands to 23 in \emph{J} and 22 in \emph{K}$_s$ (Fig.~\ref{fig:limmag}). This data set is by far the largest-area optical+near-IR data set to this depth. The galaxies in this catalogue have been detected using a reduction of the data that has been optimised for weak gravitational lensing measurements, to enable the primary science goal of KiDS. The {\sc GAaP} photometry in the nine-band catalogue uses the positions and sizes of these galaxies to define the apertures. Analysis of the gravitational lensing information in the data set is in progress, and shear estimates for these sources will be released in due course. Photometric redshift estimates based on the nine-band photometry are already included in the DR4 catalogue (but see Sect.~\ref{sec:photoz} for a discussion of redshift biases for bright sources).

Multiple other applications of this unique optical+near-infrared catalogue are foreseen. For example, stellar mass estimates for galaxies will benefit greatly from the inclusion of the near-IR fluxes \citep{wright/etal:prep}, red-sequence cluster searches can be pushed to greater redshift, and star-galaxy separation and galaxy SED typing can be made more accurate as well \citep[e.g.,][]{daddi/etal:2004,tortora/etal:2018UCMG}.

The data processing for DR4 largely followed the procedures established for the previous data release as described in \citetalias{dejong/etal:2017}, with a few improvements:
\begin{enumerate}
\item this is the first KiDS data release for which the photometry has been tied to the Gaia database;
\item satellite tracks and other artefacts are now masked at the sub-exposure level, increasing the usable area of the co-added images;
\item the PSF Gaussianization procedure now operates in pixel space, solving directly for a double-shapelet convolution kernel that renders the PSF Gaussian;
\item extinction corrections have been updated to the \cite{schlafly/finkbeiner:2011} extinction coefficients;
\item {\sc GAaP} photometry is run twice, with the second run using larger apertures to be able to include occasional poor-seeing KiDS or VIKING data in the photometry catalogue;
\item the {\sc theli} processing of the images on which the sources for the nine-band catalogue are detected now includes an illumination correction;
\item the inclusion of the VIKING data involved a re-reduction of the VIKING paw-print level data (see \citealt{wright/etal:prep}), and is the first time the PSF Gaussianization and {\sc GAaP} photometry have been performed at sub-exposure level and combined.
\end{enumerate}

The data are publicly available via the ESO archive, the {\sc Astro-WISE} system, and the KiDS project website. A description of the data format may be found in the Appendix.

KiDS observations continue, and are expected to wind down by the middle of 2019, at which point some 1350 square degrees will have been mapped in 9 photometric bands by the combined KiDS+VIKING project.

A repeat pass of the whole survey area in the \emph{i}-band is also close to completion. These data will enable variability studies on timescales of several years, as well as improving the overall quality of the \emph{i}-band data, which has the greatest variation in observing conditions and cosmetic quality. In addition, a number of fields with deep spectroscopic redshifts are also being targeted with the VST and VISTA to provide KiDS+VIKING-like photometry for large samples of faint galaxies that can be used as redshift calibrators.

The next full data release, DR5, is expected to be the final one, containing data from the full KiDS/VIKING footprint shown in Fig.~\ref{fig:fields}. Intermediate `value-added' public releases based on DR4, including one with weak lensing shape measurements, will be made together with the corresponding scientific analyses.

\begin{acknowledgements}

We are indebted to the staff at ESO-Garching and ESO-Paranal for managing the observations at VST and VISTA that yielded the data presented here.
Based on observations made with ESO Telescopes at the La Silla Paranal Observatory under programme IDs 177.A-3016, 177.A-3017, 177.A-3018 and 179.A-2004, and on data products produced by the KiDS consortium. The KiDS production team acknowledges support from: Deutsche Forschungsgemeinschaft,  ERC, NOVA and NWO-M grants; Target; the University of Padova, and the University Federico II (Naples). 
Data processing for VIKING has been contributed by the VISTA Data Flow System at CASU, Cambridge and WFAU, Edinburgh.
This work is supported by the
Deutsche Forschungsgemeinschaft in the framework of the TR33 `The Dark
Universe'. 
We acknowledge support from 
European Research Council grants 647112 (CH, BG) and 770935 (HHi),
the Deutsche Forschungsgemeinschaft
 (HHi, Emmy Noether grant Hi 1495/2-1 and Heisenberg grant Hi 1495/5-1), 
 the Polish Ministry of Science and Higher Education (MB, grant DIR/WK/2018/12),
 the INAF PRIN-SKA 2017 program 1.05.01.88.04 (CT),
 the Alexander von Humboldt Foundation (KK), the STFC (LM, grant ST/N000919/1),
and NWO (KK, JdJ, MB, HHo, research grants 621.016.402, 614.001.451 and 639.043.512). We are very grateful to the Lorentz Centre and ESO-Garching for hosting several team meetings.

Author contributions: All authors contributed to the work presented in this paper. The author list consists of three groups. The seven lead authors headed various aspects of the data processing. The second group (listed alphabetically) provided significant effort to the production of DR4. The third group contributed to the processing and quality control inspection to a lesser extent, or provided significant input to the paper.

\end{acknowledgements}

\bibliographystyle{aa}
\bibliography{KKrefs,preprints,codes}

\onecolumn
\appendix
\section{Description of the data products}
\label{app:dataproducts}

The data release consists of single-band data products, based on individual KiDS tiles observed with the \emph{u}, \emph{g}, \emph{r} and \emph{i} filters and processed with {\sc Astro-WISE}, as well as a multi-band sourcelist with photometry in 9 bands for sources detected in \emph{r}, the deepest KiDS band. The \emph{r}-band images used for this latter source list result from a separate reduction of the KiDS images, using {\sc theli}, that is optimised for weak lensing measurements. There are therefore differences between the single-band \emph{r} catalogues and the multi-band source lists. The nine-band catalogue can be queried via the ESO archive interface, or downloaded as separate fits tables per survey tile. Shape parameters for the sources in the nine-band catalogue will be released as part of the upcoming KiDS-DR4 weak lensing analysis.

File names for the various products follow the scheme given in Table~\ref{tab:filenaming}.

\begin{table}[h]
\caption{\label{tab:filenaming} File names for the data products associated with each KiDS tile. RA  and DEC  are the tile centre in degrees, to 1 decimal place.}
\label{tab:filenames}
\begin{tabular}{lll}
\hline\hline
Name & Description & Section\\
\hline
\verb KiDS_DR4.0_[RA]_[DEC]_x_sci.fits & Co-added image in band \tt x & \\
\verb KiDS_DR4.0_[RA]_[DEC]_x_wei.fits & Weight image for co-added image &\\
\verb KiDS_DR4.0_[RA]_[DEC]_x_msk.fits & Mask image for co-added image &\\
\verb KiDS_DR4.0_[RA]_[DEC]_x_src.fits & Single-band source list &\\
\verb KiDS_DR4.0_[RA]_[DEC]_r_det_sci.fits & {\sc theli} \emph{r}-band co-added image &\\
\verb KiDS_DR4.0_[RA]_[DEC]_r_det_wei.fits & {\sc theli} weight image &\\
\verb KiDS_DR4.0_[RA]_[DEC]_ugriZYJHKs_msk.fits & {\sc theli} + 9-band catalogue mask image & \\
\verb KiDS_DR4.0_[RA]_[DEC]_ugriZYJHKs_cat.fits & Nine-band catalogue &\\
\hline
\end{tabular}
\end{table}

\subsection{Single-band data products}
\subsubsection{Co-added images}
\label{app:images}
For each of the 1006 survey tiles, we provide a co-added image in each of the four KiDS filters. Image names are of the form \verb KiDS_DR4.0_RA_DEC_x_sci.fits , where \verb RA  and \verb DEC  are the nominal central coordinates of the tile, in degrees rounded to the nearest tenth, and \verb x  is the filter. Images are regridded to a grid of 0\farcs20 pixels, using a tangential projection. The images are background-subtracted. Associated inverse-variance weight images are also provided (\verb *_wei.fits ) as well as automatically generated pixel bitmasks (\verb *_msk.fits ). Different bits in the flag values indicate different types of issues: 1 = readout spike; 2 = saturation core; 4 = diffraction spike; 8 = primary reflection halo; 16 = secondary reflection halo; 32 = tertiary reflection halo; 64 = bad pixel. Several pertinent keywords that can be found in the headers of these {\sc Astro-WISE} images are listed in Table~\ref{tab:AWimkeys}.

The {\sc theli} versions of the \emph{r}-band images are also provided. These co-added images and weight file have names that contain the `\verb _det_ ' string (Table~\ref{tab:filenames}), to indicate that they are the detection images for the list-driven multi-band photometry. 

\begin{table}[h]
  \caption{Most important header keywords in the {\sc Astro-WISE} ``{\tt *sci.fits}'' images}
  \label{tab:AWimkeys}
  \begin{tabular}{ll}
    \hline\hline
    \verb KEYWORD & Description \\
    \hline
    \verb RA      &                 Field centre (J2000) (deg) \\
    \verb DEC     &                Field centre (J2000) (deg) \\
    \verb PROV[1,2,..]   & Input OmegaCAM exposures (ESO archive name)\\
    \verb ABMAGLIM  & 5-sigma limiting AB magnitude for point sources\\
    \verb ABMAGSAT  & Saturation limiting magnitude for point sources\\
    \verb PSF_FWHM  & Average seeing FWHM\\
    \verb ELLIPTIC  & Average point source ellipticity $\langle|e|\rangle$ ($1-\hbox{axis ratio})$\\
    \verb CALSTARS  & Number of stars used for Gaia+SLR calibration\\
    \verb DMAG  & Gaia+SLR photometric offset\\
    \verb CALMINAP  & \verb MIN_APER  value used for determination of \verb DMAG  \\
    \hline
  \end{tabular}
\end{table}

\subsubsection{Source lists}
\label{app:singlebandcat}
Corresponding to each single-band co-added image, DR4 provides a source list \verb KiDS_DR4.0_[RA]_[DEC]_x_src.fits , with columns as listed in Table~\ref{Tab:singlebandcolumns}. Note that the single-band catalogues include all sources detected in each observation, including ones near the edge of the field that also fall on one or more other tiles.
The structure of these catalogues is the same as it was in \citetalias{dejong/etal:2017}, but we repeat the description here for completeness. More information on some of the columns is provided below:
\begin{itemize}
\item
\verb 2DPHOT:  \textsc{KiDS-CAT} star/galaxy classification bitmap based on the source morphology \citep[see ][]{dejong/etal:2015}. Values are: 1 = high confidence star candidate; 2 = unreliable source (e.g. cosmic ray); 4 = star according to star/galaxy separation criteria; 0 = all other sources (e.g. including galaxies). Sources identified as stars can thus have a flag value of 1, 4 or 5.
\item
\verb IMAFLAGS_ISO: A bitmap of the mask flags (see Sect.~\ref{app:images}) that are set anywhere within the source's isophote.
\item
\verb FLUX_APER_*  and \verb FLUXERR_APER_* :  aperture flux measurements are included for 27 different aperture sizes. In Table~\ref{Tab:singlebandcolumns} only the smallest (2 pixels or 0\farcs4 diameter) and the largest (200 pixels or 40\arcsec\ diameter) are listed; the label for the aperture of 28.5 pixels is FLUX\_APER\_28p5.
\end{itemize}

The header of the \verb SOURCELIST  extension of these fits files contains the \verb ABMAGLIM , \verb PSF_FWHM , \verb ELLIPTIC , \verb CALSTARS , \verb DMAG  and \verb CALMINAP  keywords inherited from the corresponding science image (see Table~\ref{tab:AWimkeys}).

\begin{center}
\begin{longtable}{llll}
\caption{\label{Tab:singlebandcolumns} Columns provided in the single-band source lists.}\\
\hline\hline
Label & Format & Unit & Description \\
\hline
\endfirsthead

\multicolumn{4}{c}{\tablename\ \thetable{} -- continued from previous page (single-band source lists)}\\
\hline\hline
Label & Format & Unit & Description \\
\hline
\endhead

\hline
\multicolumn{4}{r}{{Continued on next page}}\\
\endfoot

\hline
\endlastfoot

2DPHOT & J & Source & classification (see sect.~4.5.1 in \citetalias{dejong/etal:2015})\\
X\_IMAGE & E & pixel & Object position along x      \\
Y\_IMAGE & E & pixel & Object position along y      \\
NUMBER & J &  & Running object number        \\
CLASS\_STAR & E &  & {\sc SExtractor} S/G classifier        \\
FLAGS & J &  & Extraction flags         \\
IMAFLAGS\_ISO & J &  & FLAG-image flags summed over the iso. profile\\
NIMAFLAG\_ISO & J &  & Number of flagged pixels entering IMAFLAGS\_ISO\\
FLUX\_RADIUS & E & pixel & Radius containing half of the flux \\ 
KRON\_RADIUS & E & pixel & Kron apertures in units of A or B  \\
FWHM\_IMAGE & E & pixel & FWHM assuming a gaussian core     \\
ISOAREA\_IMAGE & J & pixel$^2$ & Isophotal area above Analysis threshold     \\
ELLIPTICITY & E &  & 1 - B\_IMAGE/A\_IMAGE        \\
THETA\_IMAGE & E & deg & Position angle (CCW/x)       \\
MAG\_AUTO & E & mag & Kron-like elliptical aperture magnitude      \\
MAGERR\_AUTO & E & mag & RMS error for AUTO magnitude     \\
ALPHA\_J2000 & D & deg & Right ascension of barycenter (J2000)     \\
DELTA\_J2000 & D & deg & Declination of barycenter (J2000)      \\
FLUX\_APER\_2 & E & count & Flux within circular aperture of diameter 2 pixels  \\
... & ... & ... & ...         \\
FLUX\_APER\_200 & E & count & Flux within circular aperture of diameter 200 pixels   \\
FLUXERR\_APER\_2 & E & count & RMS error for flux within aperture of diameter 2 pixels  \\
... & ... & ... & ...         \\
FLUXERR\_APER\_200 & E & count & RMS error for flux within aperture of diameter 200 pixels  \\
MAG\_ISO & E & mag & Isophotal magnitude        \\
MAGERR\_ISO & E & mag & RMS error for isophotal magnitude     \\
MAG\_ISOCOR & E & mag & Corrected isophotal magnitude    (deprecated)   \\
MAGERR\_ISOCOR & E & mag & RMS error for corrected isophotal magnitude    \\
MAG\_BEST & E & mag & Best of MAG\_AUTO and MAG\_ISOCOR     \\
MAGERR\_BEST & E & mag & RMS error for MAG\_BEST      \\
BACKGROUND & E & count & Background at centroid position      \\
THRESHOLD & E & count & Detection threshold above background      \\
MU\_THRESHOLD & E & arcsec$^{-2}$ & Detection threshold above background      \\
FLUX\_MAX & E & count & Peak flux above background      \\
MU\_MAX & E & arcsec$^{-2}$ & Peak surface brightness above background     \\
ISOAREA\_WORLD & E & deg$^2$ & Isophotal area above Analysis threshold     \\
XMIN\_IMAGE & J & pixel & Minimum x-coordinate among detected pixels     \\
YMIN\_IMAGE & J & pixel & Minimum y-coordinate among detected pixels     \\
XMAX\_IMAGE & J & pixel & Maximum x-coordinate among detected pixels     \\
YMAX\_IMAGE & J & pixel & Maximum y-coordinate among detected pixels     \\
X\_WORLD & D & deg & Barycentre position along world x axis    \\
Y\_WORLD & D & deg & Barycentre position along world y axis    \\
XWIN\_IMAGE & E & pixel & Windowed position estimate along x     \\
YWIN\_IMAGE & E & pixel & Windowed position estimate along y     \\
X2\_IMAGE & D & pixel$^2$ & Variance along x       \\
Y2\_IMAGE & D & pixel$^2$ & Variance along y       \\
XY\_IMAGE & D & pixel$^2$ & Covariance between x and y     \\
X2\_WORLD & E & deg$^2$ & Variance along X-WORLD (alpha)      \\
Y2\_WORLD & E & deg$^2$ & Variance along Y-WORLD (delta)      \\
XY\_WORLD & E & deg$^2$ & Covariance between X-WORLD and Y-WORLD     \\
CXX\_IMAGE & E & pixel$^{-2}$ & Cxx object ellipse parameter      \\
CYY\_IMAGE & E & pixel$^{-2}$ & Cyy object ellipse parameter      \\
CXY\_IMAGE & E & pixel$^{-2}$ & Cxy object ellipse parameter      \\
CXX\_WORLD & E & deg$^{-2}$ & Cxx object ellipse parameter (WORLD units)    \\
CYY\_WORLD & E & deg$^{-2}$ & Cyy object ellipse parameter (WORLD units)    \\
CXY\_WORLD & E & deg$^{-2}$ & Cxy object ellipse parameter (WORLD units)    \\
A\_IMAGE & D & pixel & Profile RMS along major axis     \\
B\_IMAGE & D & pixel & Profile RMS along minor axis     \\
A\_WORLD & E & deg & Profile RMS along major axis (WORLD units)   \\
B\_WORLD & E & deg & Profile RMS along minor axis (WORLD units)   \\
THETA\_WORLD & E & deg & Position angle (CCW/world-x)       \\
THETA\_J2000 & E & deg & Position angle (east of north) (J2000)    \\
ELONGATION & E & deg & A\_IMAGE/B\_IMAGE         \\
ERRX2\_IMAGE & E & pixel$^2$ & RMS error on variance of position along x     \\
ERRY2\_IMAGE & E & pixel$^2$ & RMS error on variance of position along y     \\
ERRXY\_IMAGE & E & pixel$^2$ & RMS error on covariance between x and y position  \\
ERRX2\_WORLD & E & deg$^2$ & RMS error on variance of position along X-WORLD (alpha)    \\
ERRY2\_WORLD & E & deg$^2$ & RMS error on variance of position along Y-WORLD (delta)    \\
ERRXY\_WORLD & E & deg$^2$ & RMS error on covariance between X-WORLD and Y-WORLD      \\
ERRCXX\_IMAGE & E & pixel$^{-2}$ & Cxx error ellipse parameter      \\
ERRCYY\_IMAGE & E & pixel$^{-2}$ & Cyy error ellipse parameter      \\
ERRCXY\_IMAGE & E & pixel$^{-2}$ & Cxy error ellipse parameter      \\
ERRCXX\_WORLD & E & deg$^{-2}$ & Cxx error ellipse parameter (WORLD units)    \\
ERRCYY\_WORLD & E & deg$^{-2}$ & Cyy error ellipse parameter (WORLD units)    \\
ERRCXY\_WORLD & E & deg$^{-2}$ & Cxy error ellipse parameter (WORLD units)    \\
ERRA\_IMAGE & E & pixel & RMS position error along major axis    \\
ERRB\_IMAGE & E & pixel & RMS position error along minor axis    \\
ERRA\_WORLD & E & deg & World RMS position error along major axis   \\
ERRB\_WORLD & E & deg & World RMS position error along minor axis   \\
ERRTHETA\_IMAGE & E & deg & Error ellipse position angle (CCW/x)     \\
ERRTHETA\_WORLD & E & deg & Error ellipse position angle (CCW/world-x)     \\
ERRTHETA\_J2000 & E & deg & J2000 error ellipse pos. angle (east of north)  \\
FWHM\_WORLD & E & deg & FWHM assuming a gaussian core     \\
ISO0 & J & pixel$^2$ & Isophotal area at level 0     \\
ISO1 & J & pixel$^2$ & Isophotal area at level 1     \\
ISO2 & J & pixel$^2$ & Isophotal area at level 2     \\
ISO3 & J & pixel$^2$ & Isophotal area at level 3     \\
ISO4 & J & pixel$^2$ & Isophotal area at level 4     \\
ISO5 & J & pixel$^2$ & Isophotal area at level 5     \\
ISO6 & J & pixel$^2$ & Isophotal area at level 6     \\
ISO7 & J & pixel$^2$ & Isophotal area at level 7     \\
SLID & K &  & \textsc{Astro-WISE} SourceList identifier       \\
SID & K &  & \textsc{Astro-WISE} source identifier     \\
HTM & K &  & Hierarchical Triangular Mesh (level 25)      \\
FLAG & K &  & Not used         \\
\hline
\end{longtable}
\end{center}

\subsection{Nine-band catalogue}
\label{app:ninebandcat}

The production of the nine-band catalogues is described in Sect.~\ref{sec:ninebandcat}. The data release consists of 1006 catalogues one per KiDS tile. The catalogues have been cut in right ascension and declination at the edges of the fields to prevent duplication of sources in neighbouring tiles, so that the catalogues fit together seamlessly. The cuts are based on the KiDS survey pattern. Since the VIKING survey uses an instrument with a different footprint on the sky, different VIKING survey tiles generally contribute to the data for a single catalogue.

The nine-band catalogue files are called \verb KiDS_DR4.0_[RA]_[DEC]_ugriZYJHKs_cat.fits .
The columns in the nine-band catalogues are described in Table~\ref{Tab:MultiBandColumns}.
The nine-band catalogues' \verb MASK  column is a combination of the mask values of the single-band observations, and its meaning is given in Table~\ref{tab:ninebandmask}. They should not be confused with the \verb MASK  flags in the single-band catalogues.

In addition, the header of each catalogue contains keywords that reference the KiDS input data sets. The most pertinent of these are listed in Table~\ref{tab:ninebandhead}.

\begin{center}
\begin{longtable}{llll}
\caption{\label{Tab:MultiBandColumns} Columns provided in the
  nine-band catalogue.}\\
\hline\hline
Label & Unit & Format & Description \\
\hline
\endfirsthead

\multicolumn{4}{c}{\tablename\ \thetable{} -- continued from previous page (nine-band catalogue)}\\
\hline\hline
Label & Unit & Format & Description \\
\hline
\endhead

\hline
\multicolumn{4}{r}{{Continued on next page}}\\
\endfoot

\hline
\endlastfoot

\verb ID &   & 30A & ESO ID  \\
\verb SeqNr &   & 1J & Running object number within the catalogue         \\
\verb SLID &   & 1J & {\sc Astro-WISE} Source list ID           \\
\verb SID &   & 1J & {\sc Astro-WISE} Source ID within the source list        \\
\verb THELI_NAME &   & 14A & Name of the pointing in {\sc theli} convention        \\
\verb KIDS_TILE &   & 14A & Name of the pointing in {\sc Astro-WISE} convention        \\
\\
\multicolumn{4}{l}{$\qquad$Parameters derived from the {\sc theli} $r$-band detection image:}\\
\\
\verb FLUX_AUTO & count  & 1E & r-band flux             \\
\verb FLUXERR_AUTO & count  & 1E & Error on \verb FLUX_AUTO            \\
\verb MAG_AUTO & mag  & 1E & r-band magnitude            \\
\verb MAGERR_AUTO & mag  & 1E & Error on \verb MAG_AUTO           \\
\verb KRON_RADIUS & pixel& 1E & Scaling radius of the ellipse for magnitude measurements\\
\verb BackGr & count  & 1E & Background counts at centroid position         \\
\verb Level & count  & 1E & Detection threshold above background         \\
\verb MU_THRESHOLD & mag $\cdot$ arcsec$^{-2}$& 1E &Detection threshold above background\\
\verb MaxVal & count  & 1E & Peak flux above background          \\
\verb MU_MAX & mag $\cdot$ arcsec$^{-2}$  & 1E & Peak surface brightness above background\\
\verb ISOAREA_WORLD & deg$^2$  & 1E & Isophotal area above analysis threshold         \\
\verb Xpos & pixel  & 1E & Centroid x position in the {\sc theli} image       \\
\verb Ypos & pixel  & 1E & Centroid y position in the {\sc theli} image        \\
\verb RAJ2000 & deg  & 1D & Centroid sky position right ascension (J2000)        \\
\verb DECJ2000 & deg  & 1D & Centroid sky position declination (J2000)          \\
\verb A_WORLD & deg  & 1E & Profile RMS along major axis          \\
\verb B_WORLD & deg  & 1E & Profile RMS along minor axis          \\
\verb THETA_J2000 & deg  & 1E & Position angle (West of North)          \\
\verb THETA_WORLD & deg  & 1E & Position angle (Counterclockwise from world x-axis   \\
\verb ERRA_WORLD & deg  & 1E & World RMS position error along major axis        \\
\verb ERRB_WORLD & deg  & 1E & World RMS position error along minor axis        \\
\verb ERRTHETA_J2000 & deg  & 1E & Error on \verb THETA_J2000            \\
\verb ERRTHETA_WORLD & deg  & 1E & Error on \verb THETA_WORLD            \\
\verb FWHM_IMAGE & pixel  & 1E & FWHM assuming a gaussian object profile         \\
\verb FWHM_WORLD & deg  & 1E & FWHM assuming a gaussian object profile         \\
\verb Flag &   & 1I & SExtractor extraction flags            \\
\verb FLUX_RADIUS & pixel  & 1E & Half-light radius             \\
\verb CLASS_STAR &   & 1E & Star-galaxy classifier             \\
\verb MAG_ISO & mag  & 1E & r-band Isophotal Magnitude      \\
\verb MAGERR_ISO & mag  & 1E & Error on \verb MAG_ISO        \\
\verb FLUX_ISO & count  & 1E & r-band Isophotal Flux           \\
\verb FLUXERR_ISO & count  & 1E & Error on \verb FLUX_ISO            \\
\verb MAG_ISOCOR & mag  & 1E & r-band Corrected Isophotal Magnitude      \\
\verb MAGERR_ISOCOR & mag  & 1E & Error on \verb MAG_ISOCOR        \\
\verb FLUX_ISOCOR & count  & 1E & r-band Corrected Isophotal Flux       \\
\verb FLUXERR_ISOCOR & count  & 1E & Error on \verb FLUX_ISOCOR         \\
\verb NIMAFLAGS_ISO &   & 1I & Number of flagged pixels over the isophotal profile    \\
\verb IMAFLAGS_ISO &   & 1I & FLAG-image flags ORed over the isophotal profile      \\
\verb XMIN_IMAGE & pixel  & 1I & Minimum x-coordinate among detected pixels     \\
\verb YMIN_IMAGE & pixel  & 1I & Minimum y-coordinate among detected pixels     \\
\verb XMAX_IMAGE & pixel  & 1I & Maximum x-coordinate among detected pixels      \\
\verb YMAX_IMAGE & pixel  & 1I & Maximum x-coordinate among detected pixels      \\
\verb X_WORLD & deg  & 1D & Barycentre position along world x axis         \\
\verb Y_WORLD & deg  & 1D & Barycentre position along world y axis         \\
\verb X2_WORLD & deg$^2$  & 1E & Variance of position along \verb X_WORLD  (alpha)      \\
\verb Y2_WORLD & deg$^2$  & 1E & Variance of position along \verb Y_WORLD  (delta)      \\
\verb XY_WORLD & deg$^2$  & 1E & Covariance of position \verb X_WORLD,Y_WORLD       \\
\verb ERRX2_WORLD & deg$^2$  & 1E & Error on \verb X2_WORLD            \\
\verb ERRY2_WORLD & deg$^2$  & 1E & Error on \verb Y2_WORLD            \\
\verb ERRXY_WORLD & deg$^2$  & 1E & Error on \verb XY_WORLD            \\
\verb CXX_WORLD & deg$^{-2}$  & 1E & SExtractor Cxx object ellipse parameter         \\
\verb CYY_WORLD & deg$^{-2}$  & 1E & SExtractor Cyy object ellipse parameter         \\
\verb CXY_WORLD & deg$^{-2}$  & 1E & SExtractor Cxy object ellipse parameter         \\
\verb ERRCXX_WORLD & deg$^{-2}$  & 1E & Error on \verb CXX_WORLD           \\
\verb ERRCYY_WORLD & deg$^{-2}$  & 1E & Error on \verb CYY_WORLD           \\
\verb ERRCXY_WORLD & deg$^{-2}$  & 1E & Error on \verb CXY_WORLD           \\
\verb A_IMAGE & pixel  & 1D & Profile RMS along x-axis          \\
\verb B_IMAGE & pixel  & 1D & Profile RMS along y-axis          \\
\verb ERRA_IMAGE & pixel  & 1E & Error on \verb A_IMAGE            \\
\verb ERRB_IMAGE & pixel  & 1E & Error on \verb B_IMAGE            \\
\verb S_ELLIPTICITY &   & 1E & SExtractor Ellipticity \verb (1-B_IMAGE/A_IMAGE)           \\
\verb S_ELONGATION &   & 1E & SExtractor Elongation \verb (A_IMAGE/B_IMAGE)           \\
\verb MAG_APER_4 & mag  & 1E & r-band Magnitude within a circular aperture of 4 pixels \\
\verb MAGERR_APER_4 & mag  & 1E & Error on \verb MAG_APER_4           \\
\verb FLUX_APER_4 & count  & 1E & r-band Flux within a circular aperture of 4 pixels  \\
\verb FLUXERR_APER_4 & count  & 1E & Error on \verb FLUX_APER_4         \\
\multicolumn{4}{l}{\ldots}\\
\multicolumn{4}{l}{$\qquad$Similarly for radii 6, 8, 10, 14, 20, 30, 40, 60 pixels, up to}\\
\multicolumn{4}{l}{\ldots}\\
\verb MAG_APER_100 & mag  & 1E & r-band Magnitude within a circular aperture of 100 pixels \\
\verb MAGERR_APER_100 & mag  & 1E & Error on \verb MAG_APER_100           \\
\verb FLUX_APER_100 & count  & 1E & r-band Flux within a circular aperture of 100 pixels  \\
\verb FLUXERR_APER_100 & count  & 1E & Error on \verb FLUX_APER_100         \\
\verb ISO0 & pixel$^2$  & 1I & Isophotal area at level 0         \\
\verb ISO1 & pixel$^2$  & 1I & Isophotal area at level 1         \\
\verb ISO2 & pixel$^2$  & 1I & Isophotal area at level 2          \\
\verb ISO3 & pixel$^2$  & 1I & Isophotal area at level 3         \\
\verb ISO4 & pixel$^2$  & 1I & Isophotal area at level 4         \\
\verb ISO5 & pixel$^2$  & 1I & Isophotal area at level 5          \\
\verb ISO6 & pixel$^2$  & 1I & Isophotal area at level 6       \\
\verb ISO7 & pixel$^2$  & 1I & Isophotal area at level 7       \\
\verb ALPHA_J2000 & deg  & 1D & SExtractor named Centroid sky position right ascension (J2000)\\
\verb DELTA_J2000 & deg  & 1D & SEXtractor named Centroid sky position declination (J2000) \\
\verb SG2DPHOT &   & 1I & 2DPhot StarGalaxy classifier (1 for high confidence star)    \\
\verb HTM &   & 1J & Hierarchical Triangular Mesh (level 25)          \\
\verb FIELD_POS &   & 1I & Reference number to field parameters          \\
\\
\multicolumn{4}{l}{$\qquad$List-driven {\sc GAaP} photometry on the {\sc Astro-WISE} co-added KiDS images and the pawprint VIKING images:}\\
\\
\verb Agaper_0p7 & arcsec  & 1E & Major axis of {\sc GAaP} aperture \verb MIN_APER  0.7\arcsec       \\
\verb Bgaper_0p7 & arcsec  & 1E & Minor axis of {\sc GAaP} aperture \verb MIN_APER  0.7\arcsec       \\
\verb Agaper_1p0 & arcsec  & 1E & Major axis of {\sc GAaP} aperture \verb MIN_APER  1.0\arcsec       \\
\verb Bgaper_1p0 & arcsec  & 1E & Minor axis of {\sc GAaP} aperture \verb MIN_APER  1.0\arcsec       \\
\verb PAgaap & deg  & 1E & Position angle of major axis of {\sc GAaP} aperture (North of West)\\
\\
\multicolumn{4}{l}{$\qquad\qquad$and then for each band {\tt x} = {\tt u,g,r,i,Z,Y,J,H,Ks}:\tablefootmark{a}}\\
\\
\verb FLUX_GAAP_0p7_x & count  & 1E & {\sc GAaP} flux in band \verb x  with \verb MIN_APER =0.7\arcsec         \\
\verb FLUXERR_GAAP_0p7_x & count  & 1E & Error on \verb FLUX_GAAP_0p7_x     \\
\verb MAG_GAAP_0p7_x & mag  & 1E & \verb x -band {\sc GAaP} magnitude with \verb MIN_APER =0.7\arcsec        \\
\verb MAGERR_GAAP_0p7_x & mag  & 1E & Error on \verb MAG_GAAP_0p7_x \\
\verb FLAG_GAAP_0p7_x &   & 1J & {\sc GAaP} Flag for \verb x -band photometry with \verb MIN_APER =0.7\arcsec \\
\verb FLUX_GAAP_1p0_x & count  & 1E & {\sc GAaP} flux in band \verb x  with \verb MIN_APER =1.0\arcsec        \\
\verb FLUXERR_GAAP_1p0_x & count  & 1E & Error on \verb FLUX_GAAP_1p0_x    \\
\verb MAG_GAAP_1p0_x & mag  & 1E & \verb x -band {\sc GAaP} magnitude with \verb MIN_APER =1.0\arcsec     \\
\verb MAGERR_GAAP_1p0_x & mag  & 1E & Error on \verb MAG_GAAP_1p0_x \\
\verb FLAG_GAAP_1p0_x &   & 1J & {\sc GAaP} Flag for \verb x -band photometry with \verb MIN_APER =1.0\arcsec  \\
\\
\multicolumn{4}{l}{$\qquad$Optimal-aperture {\sc GAaP} 9-band photometry including interstellar extinction corrections}\\
\\
\verb Agaper & arcsec  & 1E & Major axis of {\sc GAaP} aperture for optimal \verb MIN_APER  \\
\verb Bgaper & arcsec  & 1E & Minor axis of {\sc GAaP} aperture for optimal \verb MIN_APER  \\
\\
\multicolumn{4}{l}{$\qquad\qquad$and then for each band {\tt x}= {\tt u,g,r,i,Z,Y,J,H,Ks}:\tablefootmark{a}}\\
\\
\verb EXTINCTION_x & mag  & 1E & Galactic extinction in band \verb x         \\
\verb MAG_GAAP_x & mag  & 1E & \verb x -band {\sc GAaP} magnitude with optimal \verb MIN_APER  (extinction corrected)\\
\verb MAGERR_GAAP_x & mag  & 1E & Error on \verb MAG_GAAP_x \\
\verb FLUX_GAAP_x & count  & 1E & \verb x -band {\sc GAaP} flux with optimal \verb MIN_APER           \\
\verb FLUXERR_GAAP_x & count  & 1E & Error on \verb FLUX_GAAP_x \\
\verb FLAG_GAAP_x &   & 1I & {\sc GAaP} Flag for \verb x band photometry with optimal \verb MIN_APER \\
\verb MAG_LIM_x  & mag   & 1E & \verb x -band limiting magnitude for optimal \verb MIN_APER \\
\\
\multicolumn{4}{l}{$\qquad$9-band photometric redshifts (BPZ):}
\\
\verb Z_B &   & 1D & 9-band BPZ redshift estimate; peak of posterior probability distribution      \\
\verb Z_B_MIN &   & 1D & Lower bound of the 68\% confidence interval of \verb Z_B      \\
\verb Z_B_MAX &   & 1D & Upper bound of the 68\% confidence interval of \verb Z_B      \\
\verb T_B &   & 1D & Spectral type corresponding to \verb Z_B          \\
\verb ODDS &   & 1D & Empirical ODDS of \verb Z_B            \\
\verb Z_ML &   & 1D & 9-band BPZ maximum likelihood redshift          \\
\verb T_ML &   & 1D & Spectral type corresponding to \verb Z_ML \tablefootmark{b}         \\
\verb CHI_SQUARED_BPZ &   & 1D & chi sqared value associated with \verb Z_B         \\
\verb M_0 & mag  & 1D & Reference magnitude for BPZ prior          \\
\verb BPZ_FILT &   & 1J & filters with good photometry (BPZ)          \\
\verb NBPZ_FILT &   & 1J & number of filters with good photometry (BPZ)        \\
\verb BPZ_NONDETFILT &   & 1J & filters with faint photometry (BPZ)          \\
\verb NBPZ_NONDETFILT &   & 1J & number of filters with faint photometry (BPZ)        \\
\verb BPZ_FLAGFILT &   & 1J & flagged filters (BPZ)            \\
\verb NBPZ_FLAGFILT &   & 1J & number of flagged filters (BPZ) \\
\verb SG_FLAG &   & 1E & Star/Gal Classifier\\
\verb MASK &   & 1J & 9-band mask information \tablefootmark{c}           \\
\\
\multicolumn{4}{l}{$\qquad$Dereddened colours based on optimal-aperture {\sc GAaP} photometry}\\
\\
\verb COLOUR_GAAP_u_g & mag  & 1E & u-g colour index  (dereddened)          \\
\verb COLOUR_GAAP_g_r & mag  & 1E & g-r colour index  (dereddened)          \\
\verb COLOUR_GAAP_r_i & mag  & 1E & r-i colour index  (dereddened)          \\
\verb COLOUR_GAAP_i_Z & mag  & 1E & i-Z colour index   (dereddened)         \\
\verb COLOUR_GAAP_Z_Y & mag  & 1E & Z-Y colour index   (dereddened)         \\
\verb COLOUR_GAAP_Y_J & mag  & 1E & Y-J colour index   (dereddened)         \\
\verb COLOUR_GAAP_J_H & mag  & 1E & J-H colour index  (dereddened)          \\
\verb COLOUR_GAAP_H_Ks & mag  & 1E & H-Ks colour index   (dereddened)         \\

\hline
\end{longtable}
\tablefoot{
\tablefoottext{a}{See \S\ref{sec:ninebandcat} for the definitions of the flux and magnitude zeropoints.}
  \tablefoottext{b}{Definition of the spectral types: 1=CWW-Ell, 2=CWW-Sbc, 3=CWW-Scd, 4=CWW-Im,
5=KIN-SB3, 6=KIN-SB2}
\tablefoottext{c}{For the meaning of the mask see Table~\ref{tab:ninebandmask}}
}

\end{center}

\begin{table}[h]
\caption{Bit values of the nine-band {\tt MASK} parameter}
\label{tab:ninebandmask}
\begin{tabular}{ll}
\hline\hline
Bit & Meaning \\
\hline
0 & {\sc theli} manual mask (very conservative)\\
1 & {\sc theli} automatic large star halo mask (faint)\tablefootmark{a}\\
2 & {\sc theli} automatic large star halo mask (bright) or star mask\tablefootmark{b}\\
3 & manual mask of regions around globular clusters, Fornax dwarf, ISS\\
4 & {\sc theli} weight=0, or void mask, or asteroids\\
5 & VIKING $Z$ band masked\tablefootmark{c}\\
6 & VIKING $Y$ band masked\\
7 & VIKING $J$ band masked\\
8 & VIKING $H$ band masked\\
9 & VIKING $K_{\rm s}$ band masked\\
10 &{\sc Astro-WISE} $u$ band halo+stellar {\sc pulecenella} mask or weight=0\\
11 &{\sc Astro-WISE} $g$ band halo+stellar {\sc pulecenella} mask or weight=0\\
12 &{\sc Astro-WISE} $r$ band halo+stellar {\sc pulecenella} mask\tablefootmark{d} or weight=0\\
13 &{\sc Astro-WISE} $i$ band halo+stellar {\sc pulecenella} mask or weight=0\\
14 & Object outside the RA/DEC cut for its tile\\
15 & Reserved (used for sign in FITS 2-byte integer)\\
\hline
\end{tabular}
\tablefoot{
\tablefoottext{a}{Haloes from stars with $10.5<m_r<11.5$ in UCAC4/GSC1 stellar catalog.}
\tablefoottext{b}{Stars with $m_r<14$ or haloes from stars with $m_r<10.5$ in UCAC4/GSC1.}
\tablefoottext{c}{The VIKING masks are described in \cite{wright/etal:prep}.}
\tablefoottext{d}{Note that these tend to more conservative than the {\sc theli} flag in bit 2.}
}
\end{table}

\begin{table}[h]
\caption{Main keywords in the nine-band catalogue headers}
\label{tab:ninebandhead}
\begin{tabular}{ll}
\hline\hline
Keyword & Description \\
\hline
\verb RA      &                 Field centre (J2000) (deg) \\
\verb DEC     &                Field centre (J2000) (deg) \\
\verb PROV[1,2,3,4]   & Originating [\emph{u,g,r,i}]-band co-add file name\\
\verb FPRA[1,2,3,4]   &         Footprint [SE,NE,NW,SW] corner RA (deg) \\
\verb FPDE[1,2,3,4]   &       Footprint [SE,NE,NW,SW] corner DEC (deg) \\
\verb CALSTARS &   Number of stars used for Gaia calibration (\verb MIN_APER =0\farcs7) \\
\verb D[U,G,R,I]_SLR  &    SLR [\emph{u,g,r,i}]-band offset  (\verb MIN_APER =0\farcs7) \\
\verb DMAG_[U,G,R,I]  &   SLR+Gaia  [\emph{u,g,r,i}]-band offset  (\verb MIN_APER =0\farcs7) \\
\verb CALSTR_1 & Number of stars used for Gaia calibration (\verb MIN_APER =1\farcs0) \\
\verb D[U,G,R,I]_SLR_1 & SLR  [\emph{u,g,r,i}]-band offset  (\verb MIN_APER =1\farcs0) \\
\verb DMAG_[U,G,R,I]_1 &  SLR+Gaia  [\emph{u,g,r,i}]-band offset  (\verb MIN_APER =1\farcs0) \\
\verb OB[U,G,R,I]_STRT &  [\emph{u,g,r,i}]-band Observing Block start \\
\verb ASSON1  & Associated nine-band mask\\
\verb ASSON2  & Associated {\sc theli} \emph{r}-band detection image \\
\verb ASSON3  & Associated {\sc theli} \emph{r}-band weightimage \\
\hline
\end{tabular}
\end{table}

\end{document}